\documentclass[10pt,a4paper,twocolumn,english,prb,aps,floatfix,groupedaddress,superscriptaddress]{revtex4-2}

\usepackage{graphicx}
\usepackage{epsfig}
\usepackage[english]{babel}
\usepackage{amsmath}
\usepackage{amssymb}
\usepackage{amsfonts}
\usepackage{dcolumn}
\usepackage{bm}
\usepackage{bbm}
\usepackage{nicefrac}
\usepackage{color,array}
\usepackage{colortbl}
\usepackage{hyperref}
\usepackage[caption=false,position=top]{subfig}
\usepackage{dsfont}
\usepackage{braket}
\usepackage{mathtools}

\def\Bext{B_\mathrm{ext}}
\def\Bov{\B_\mathrm{Ov}}
\def\Beff{B_\mathrm{eff}}
\def\Omegaeff{\Omega_\mathrm{eff}}
\def\Bdnp{B_\mathrm{DNP}}
\def\ddt{\frac{d}{dt}}
\def\ex{\bm{e}_x}
\def\ey{\bm{e}_y}
\def\ez{\bm{e}_z}
\def\S{\bm{S}}
\def\J{\bm{J}}
\def\B{\bm{B}}
\def\Q{\bm{Q}}
\def\Ntr{N_\mathrm{tr}}
\def\Neff{N_\mathrm{eff}}
\def\TR{T_\mathrm{R}}
\def\Tnstar{T_\mathrm{n}^*}
\def\T2star{T_\mathrm{2}^*}

\def\np{n_\mathrm{p}}
\def\ge{g_\mathrm{e}}
\def\gh{g_\mathrm{h}}
\def\gn{g_\mathrm{n}}
\def\muB{\mu_\mathrm{B}}
\def\mun{\mu_\mathrm{n}}
\def\Slim{S^\perp_\mathrm{lim}}

\def\AQ{A_\mathrm{Q}}
\def\SSML{S_\mathrm{SML}}

\begin{document}

\title{Interplay of spin mode locking and nuclei-induced frequency focusing in quantum dots}

\author{Philipp Schering}%
\email{philipp.schering@tu-dortmund.de}
\author{Philipp W. Scherer}
\author{G\"otz S. Uhrig}%
\email{goetz.uhrig@tu-dortmund.de}
\affiliation{%
	Lehrstuhl f\"ur Theoretische Physik I, TU Dortmund University, 44221 Dortmund, Germany
}%

\date{\today}

\begin{abstract}
We study the influence of nuclei-induced frequency focusing on the mode locking of spin coherence in quantum dots subjected to a periodic train of optical pulses. In particular, we address the question whether or not nuclei-induced frequency focusing always enhances the effect of spin mode locking. We combine two advanced semiclassical approaches and extend the resulting model by including the full dynamics of the optically excited trion state. In order to reduce the discrepancy to a full quantum model, we establish a nondeterministic pulse description by interpreting each pump pulse as a measurement. Both extensions lead to significant qualitative changes of the physics. Their combination improves the description of the corresponding experiments. Importantly, we observe the emergence of dynamic nuclear polarization, i.e., the formation of a nonzero average polarization of the nuclear spin bath, leading to a certain increase of the coherence time.
\end{abstract}

\maketitle

\section{Introduction}
\label{sec:introduction}

The generation of well-controllable stable quantum states is an ever ongoing challenge in the context of quantum information.
A promising candidate for a technological realization is an electron spin localized in a semiconductor quantum dot~(QD)~\cite{divincenzo98} due to the established fabrication tools for semiconductor nanostructures and the possible scalability.
The major problem is the interaction of the quantum states with the environment, eventually leading to decoherence.

Recently, important progress has been made in this field by \citet{gangloff19}, who demonstrated the implementation and manipulation of coherent states in a nuclear spin ensemble coupled to a localized electron spin in a QD, which they achieved by exploiting the hyperfine interaction with the surrounding nuclei.
This step can be seen as the ``missing piece of the puzzle'' for a semiconductor nanostructure platform for quantum information~\cite{bayer19}.

In a related earlier experiment by \citet{greil07a} on QD ensembles in a transverse external magnetic field (Voigt geometry) subjected to trains of optical pulses, it was demonstrated that the hyperfine interaction can be also exploited such that the nuclear spin bath acts as a correction field to the Zeeman term which varies from QD to QD.
By means of the nuclei-induced frequency focusing~(NIFF) effect, the nuclear spin bath is manipulated in such a way that the Larmor precession of the localized electron spins is focused onto very few resonances, enhancing the effect of spin mode locking~(SML)~\cite{greil06b}.

The SML effect is briefly described as follows.
Usually, the optically induced polarization of the localized electron spins dephases quickly due to a broad and inhomogeneous spectrum of precession frequencies.
By applying trains of periodic pulses to the QD ensemble, a revival of the spin polarization emerges before the arrival of the next pulse. 
The amplitude of these revivals, depending on many system parameters, can be strongly enhanced by a frequency focusing in the nuclear spin bath, which is achieved by applying a very long train of periodical pulses to the system.
The typical pulse repetition time is $\TR = 13.2$~ns, applied as a train for up to a few minutes so that a comprehensive theoretical description is a tremendous challenge. For the simulation of a realistic experimental setup, one has to cover about 12 orders of magnitude in time.

Through optimization of the experimental protocol, it is even possible to drive the spectrum of Larmor frequencies to only a narrow single mode such that all localized electron spins precess with almost the same frequency~\cite{greil09}.
Effectively, this leads to a strong increase of the coherence time and thus, enables the coherent manipulation of the localized electron spins~\cite{greil09b}.

In the context of SML and NIFF, there are several open questions stemming from recent experiments~\cite{varwig14,jasch17,klein18,evers18}.
A fundamental one is whether or not NIFF always acts constructively, i.e., does it always lead to an enhancement of the SML effect, and what is the influence of the external magnetic field strength on this interplay?
Recent theoretical studies, both quantum mechanical~\cite{beuge17,klein18} and semiclassical~\cite{jasch17}, suggest that this is not necessarily the case due to the possibility of additional resonances which can act destructively.
The slow Larmor precession of the nuclear spins plays a major role in this context.
In these studies, however, the full dynamics of the excited trion state is neglected. 
We lift this simplification in the present paper and demonstrate that this extension influences the physics qualitatively.

The present paper is devoted to a better theoretical description and understanding of NIFF.
The existing semiclassical precession models~\cite{glazo12a,petrov12,jasch17,scher18,klein18} are improved by interpreting each optical pump pulse as a measurement~\cite{scher18,klein18}.
This allows us to apply a truncated Wigner approximation~\cite{polkovnikov10} to the action of each pulse, leading to a reduced discrepancy to a full quantum mechanical description.
Moreover, we investigate the role of the dynamics of the optically excited trion state and the role of an inhomogeneous ensemble of QDs.
Quantum models describing the effect of NIFF also exist~\cite{greil07a,carter09,barnes11a,petrov12,economou14,beuge16,beuge17,klein18}, but they are typically restricted to a very small number of nuclear spins due to the exponentially growing Hilbert space so that one needs to either resort to a constant distribution of the hyperfine couplings (box model) or to perturbation theory.

In the following section, we introduce the initial model for the description of a homogeneous ensemble of QDs in a transverse magnetic field subjected to periodic circularly polarized laser pulses, which is a combination of various approaches from the literature~\cite{yugov09,jasch17,fauseweh17,scher18}.
However, we will show that this model does not describe the experimental results appropriately.
We extend this initial model step by step in the subsequent sections, leading to the extended models (EM) I, II, and III.
A nondeterministic description of the pulse model is introduced in Sec.~\ref{sec:EM_I} by interpreting each pulse as a measurement. 
This reduces the discrepancy to a full quantum model while being able to treat large numbers of nuclear spins for a realistic distribution of the hyperfine couplings.
In Sec.~\ref{sec:EM_II}, we extend the model by including the full trion dynamics, leading to qualitatively different physics and importantly, to the emergence of dynamic nuclear polarization, i.e., a finite average polarization in the nuclear spin bath.
Here, we emphasize the need for further experimental studies with explicit suggestions.
The role of inhomogeneities in the QD ensemble is briefly discussed in Sec.~\ref{sec:EM_III}.
Finally, a conclusion and an outlook are given in Sec.~\ref{sec:conclusion}.

\section{Initial model:\\Localized electron spin in a quantum~dot subjected to periodic optical~pulses}
\label{sec:initial_model}

In this section, we introduce and numerically analyze the initial model in which we combine an established pulse model often used to describe the excitation of a trion~\cite{greil06a,yugov09,yugov12,jasch17,jasch18,scher19} with an efficient approach to the dynamics of the nuclear spin bath~\cite{fauseweh17,scher18}, which allows us to simulate the saturation behavior of the system after very long trains of periodic optical pulses.
Note that this model does not yield an appropriate description of some experimental results, but it is a good starting point to introduce the basic phenomena of SML and NIFF.

\subsection{Equations of motion}

\begin{figure}[t!]
	\centering
	\includegraphics[width=\columnwidth]{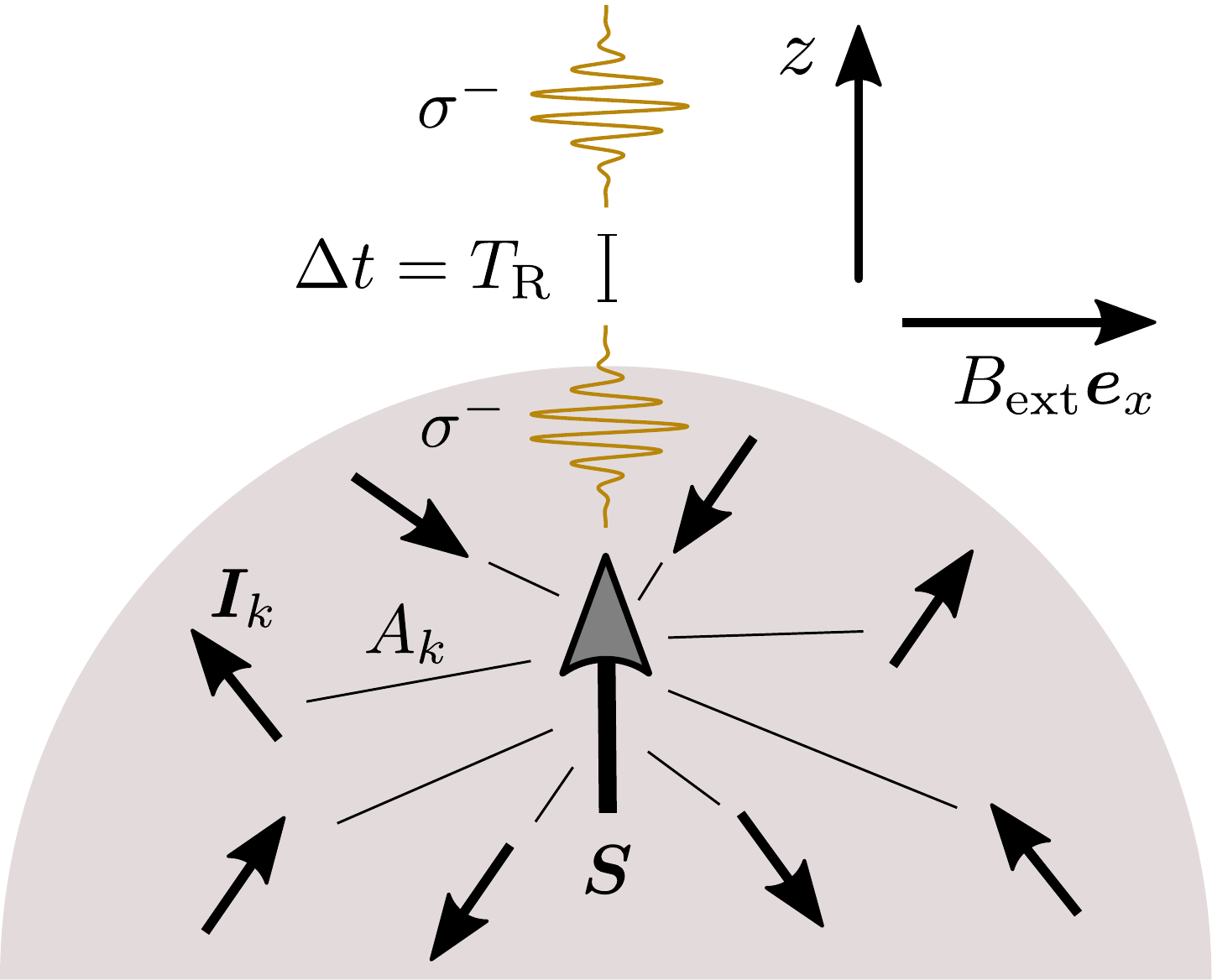}
	\caption{Sketch of the model: A localized electron spin~$\S$ in a quantum dot is subjected to a train of periodic $\sigma^-$~pulses with repetition time~$\TR$. The spin $\S$ couples to the surrounding nuclear spins~$\bm I_k$ via the hyperfine interaction with coupling constants~$A_k$. An external magnetic field~$\Bext \ex$ is applied in Voigt geometry, i.e., perpendicular to the growth direction~$\ez$, which is also the axis of incidence of the laser beam.
	}	\label{fig:sketch}
\end{figure}

We consider a homogeneous ensemble of GaAs QDs, i.e., all QDs are equal. 
They are singly charged by electrons and subjected to a strong transverse external magnetic field (Voigt geometry) of up to several Tesla.
A sketch which depicts the basic model and setup is shown in Fig.~\ref{fig:sketch}.
In each QD, the internal spin dynamics are governed by the hyperfine interaction of the single localized electron spin $\hat{\S}$ with the surrounding $N$ nuclear spins $\hat{\bm I}_k$ of the host lattice~\cite{merku02,khaetskii02,schliemann03,coish04,erlingsson04,braun05,hanson07,urba13}.
Quantum mechanically, this interaction is described by the hyperfine Hamiltonian
\begin{align}
	\hat{\mathcal{H}}_\mathrm{HF} = \sum_{k=1}^N A_k \hat{\S} \cdot \hat{\bm I}_k = \hat{\S} \cdot \hat{\bm B}_\mathrm{Ov},
\end{align}
with the hyperfine coupling constants $A_k$ together with the nuclear spins $\hat{\bm  I}_k$ forming the so called Overhauser field 
\begin{align}
	\hat{\bm B}_\mathrm{Ov} = \sum_{k=1}^N A_k \hat{\bm I}_k.
\end{align}
This Hamiltonian, often referred to as the central spin or Gaudin model~\cite{gaudin76,gaudin83,bortz07,bortz10,faribault13a,
faribault13b,alhassanieh06,chen07,barnes11b,stanek13,stanek14,uhrig14,seifert16,froehling17,roehr18,lindoy18,claeys18,froehling18,froehling19}, is extremely hard to solve for a nonuniform distribution of the couplings $A_k$ so that one is typically restricted by either the bath size $N$ or the maximum simulation time in spite of the existence of a Bethe ansatz solution~\cite{gaudin76,gaudin83,bortz07,bortz10,faribault13a,faribault13b}.
This Hamiltonian is not only of interest in the context of quantum dots, but is also used to describe radical pair recombination reactions~\cite{schulten78,manolopoulos13,lindoy20}. 
In both cases, semiclassical approaches to the spin dynamics of the system work remarkably well~\cite{stanek14,lindoy20}.

We also treat the dynamics of the system in a semiclassical manner, i.e., we solve the corresponding classical equations of motion and average over an appropriate distribution for the initial conditions of the classical spins~\cite{stanek14}.
This corresponds to a truncated Wigner approximation~\cite{polkovnikov10}, which is a semiclassical phase space method.
Because a QD comprises $N = 10^4 - 10^6$ nuclear spins~\cite{merku02,lee05,urba13}, it is well justified to consider the Overhauser field as a classical variable~$\hat{\bm B}_\mathrm{Ov} \to \Bov$ chosen randomly from a normal distribution~\cite{schulten78,merku02,chen07,stanek14}.
Since the temperature in experiments corresponds to a much larger energy than the hyperfine couplings, the nuclear spins are in a completely disordered state. 
Thus, each component $B_\mathrm{Ov}^\alpha$, $\alpha \in \{x,y,z\}$, is initially sampled from a normal distribution characterized by the expectation value $\mathrm{E}[B_\mathrm{Ov}^\alpha] = 0$ and the variance $\mathrm{Var}[B_\mathrm{Ov}^\alpha] = 2/(\Tnstar)^2$.
Unless stated otherwise, we choose a typical value of $\Tnstar = 1$~ns~\cite{greil06a,greil06b}. 
Note that we define the variance of the Overhauser field via the dephasing time $\Tnstar$ because this time is accessible in the related experiments. 
Physically it is defined via the strength of the hyperfine interaction and the spin quantum number of the nuclei; see Appendix~\ref{app:overhauser} for details.

In the semiclassical picture, the dynamics of the localized electron spin $\S$ for a single random initial configuration of the full ensemble is determined by the classical equations of motion ($\hbar$ is set to unity)
\begin{subequations}
\begin{align}
	\ddt \S &= \left( \Bov + \ge \muB \Bext \ex \right) \times \S + \frac{1}{\tau_0} J^z \ez, \label{eq:eom_S}\\
	\ddt J^z &= - \frac{1}{\tau_0} J^z, \label{eq:eom_Jz}\\
	\ddt \bm I_k &= \left(A_k \S + \gn \mun \Bext \ex \right) \times \bm I_k, \label{eq:eom_Ik},
\end{align}
\label{eq:eom}%
\end{subequations}%
$k \in \{1,2,\dots,N\}$, where $\ge = 0.555$~\cite{greil07a} is the $g$ factor of the localized electron spin, $\muB$ the Bohr magneton, $\Bext$ the strength of the external magnetic field, and $\bm{e}_\alpha$ the unit vector in $\alpha$ direction. The intermediate trion state $J^z$, which is excited by a pump pulse, has the lifetime $\tau_0 = 400$~ps~\cite{greil06a,greil06b} and eventually decays radiatively into the ground state $\S$.
The derivation of the recombination dynamics for the trion and ground state spin polarizations follows, e.g., from a Lindblad approach developed by \citet{jasch17}.
It also appears in several other works, e.g., in Refs.~\cite{shaba03,greil06a,yugov09,yugov12,smirnov18,scher19}.

Importantly, the Overhauser field $\Bov$ is a dynamic object since the nuclear spins $\bm I_k$ are also dynamic as described by Eq.~\eqref{eq:eom_Ik}.
The hyperfine coupling constants~$A_k$ are proportional to the probability of presence of the localized electron at the position of the $k$th nucleus.
For two-dimensional and flat QDs, the envelope wave function of the electron in its orbital ground state is a Gaussian envelope in a two-dimensional plane. 
It can be shown that this results in an exponential parametrization of the hyperfine couplings~\cite{coish04,fauseweh17},
\begin{align}
	A_k \propto \exp(-k\gamma),
	\label{eq:couplings}%
\end{align}
$k \in \{1,2,\dots,N\}$, which we apply in the model.
The parameter $\gamma$ defines the number of effectively coupled nuclear spins via $N_\mathrm{eff} \approx 2/\gamma$~\cite{fauseweh17,scher18,roehr18}.
Since it is unfeasible to solve the corresponding equations of motion~\eqref{eq:eom_Ik} for each individual nuclear spin $\bm I_k$ for a realistic bath size $N$, we resort to a more efficient approach.

By applying the spectral density~(SD) approach to the Overhauser field dynamics~\cite{fauseweh17}, we reduce the number of dynamic variables from $3N+4$ to $3\Ntr+4$, where ${\Ntr = \mathcal{O}(75) \ll N}$ is a truncation parameter.
The essence of this approach is the replacement of the nuclear spins $\bm I_k$ by appropriate sums of nuclear spins, represented by the auxiliary vectors $\Q_k$, similar to an approach by \citet{erlingsson04}.
These vectors follow the equation of motion
\begin{align}
	\ddt \Q_k = \left(\epsilon_k  \S + \gn \mun \Bext \ex \right) \times \Q_k,
	\label{eq:Qk}
\end{align}
$k \in \{1,2,\dots,\Ntr\}$, where $\gn \mun = \ge \muB / 800$ is roughly the average magnetic moment of the nuclei in a GaAs QD~\cite{coish09,beuge17,jasch17,scher18,klein18}, which is roughly $800$~times smaller than the magnetic moment of the localized electron due to the much larger masses of the nuclei, and the $\epsilon_k \propto \sqrt{2\gamma}/\Tnstar$ are effective coupling constants which emerge from the original couplings distribution~\eqref{eq:couplings} via application of the SD approach; see Appendix~\ref{app:SDA}.
Finally, the Overhauser field is given by 
\begin{align}
	\Bov = \sum_{i=1}^{\Ntr} \sqrt{W_k} \Q_k, \label{eq:Overhauser_SDA}
\end{align}
where the $W_k$ are weights also emerging from the SD approach. 
Since we focus on GaAs QDs, we choose $I = 3/2$ for the nuclear spins. 
Thus, we sample the components of the vectors $\Q_k$ from a normal distribution around zero with variance $I(I+1)/3 = 5/4$.
For InGaAs QDs, one needs to also account for $I = 9/2$ of the indium isotopes.

A detailed explanation of the SD approach is given in Appendix~\ref{app:SDA} and in Refs.~\cite{fauseweh17,scher18}.

\subsection{Pulse model}

The periodic optical pumping is carried out with the repetition time~$\TR = 13.2$~ns~\cite{greil06a,greil06b,greil07a,klein18}.
We focus on resonant pumping of the electron spin $\S$ by circularly polarized $\pi$ pulses with helicity $\sigma^-$, which have a typical duration of $1.5$~ps in the experiments~\cite{greil06a,greil06b,greil07a,greil07c,greil09b,evers18,klein18}.
The pumping of the localized electron spin with this circularly polarized light leads to the excitation of a negatively charged singlet trion $X^-$, which decays completely before the arrival of the next pulse under the experimental condition $\tau_0 \ll \TR$.
The trion consists of two electrons in a spin singlet state and a heavy hole with unpaired spin.
We consider flat QDs where the lateral size by far exceeds their height, i.e., we can choose the growth axis $\ez$ to be the quantization axis for the electron and heavy-hole spin states.
For $\sigma^-$~pulses, the electron spin state $-1/2$ and the heavy hole spin state $-3/2$ are responsible for the dominant optical transition due to the conservation of angular momentum~\cite{yugov09,glazo12b,urba13}.
Because one Larmor period even in a magnetic field as large as $9$~T lasts about $14$~ps, we consider the pulse to act instantaneously since the pulse duration is shorter by one order of magnitude.

\begin{figure*}[t!]
	\centering
	\subfloat{\includegraphics[width=\columnwidth]{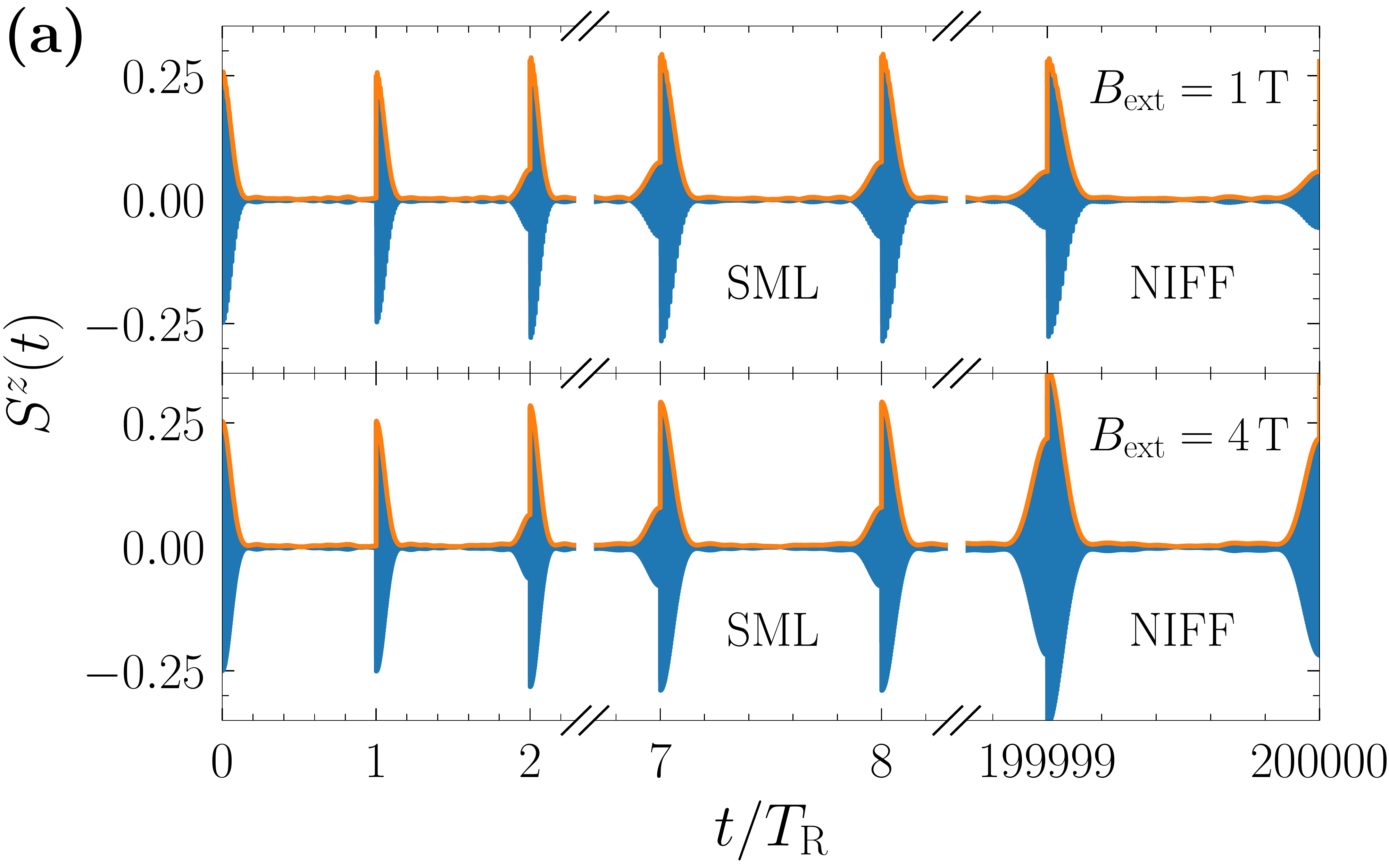} \label{fig:TD_S_timeevolution}}	
	\subfloat{\includegraphics[width=\columnwidth]{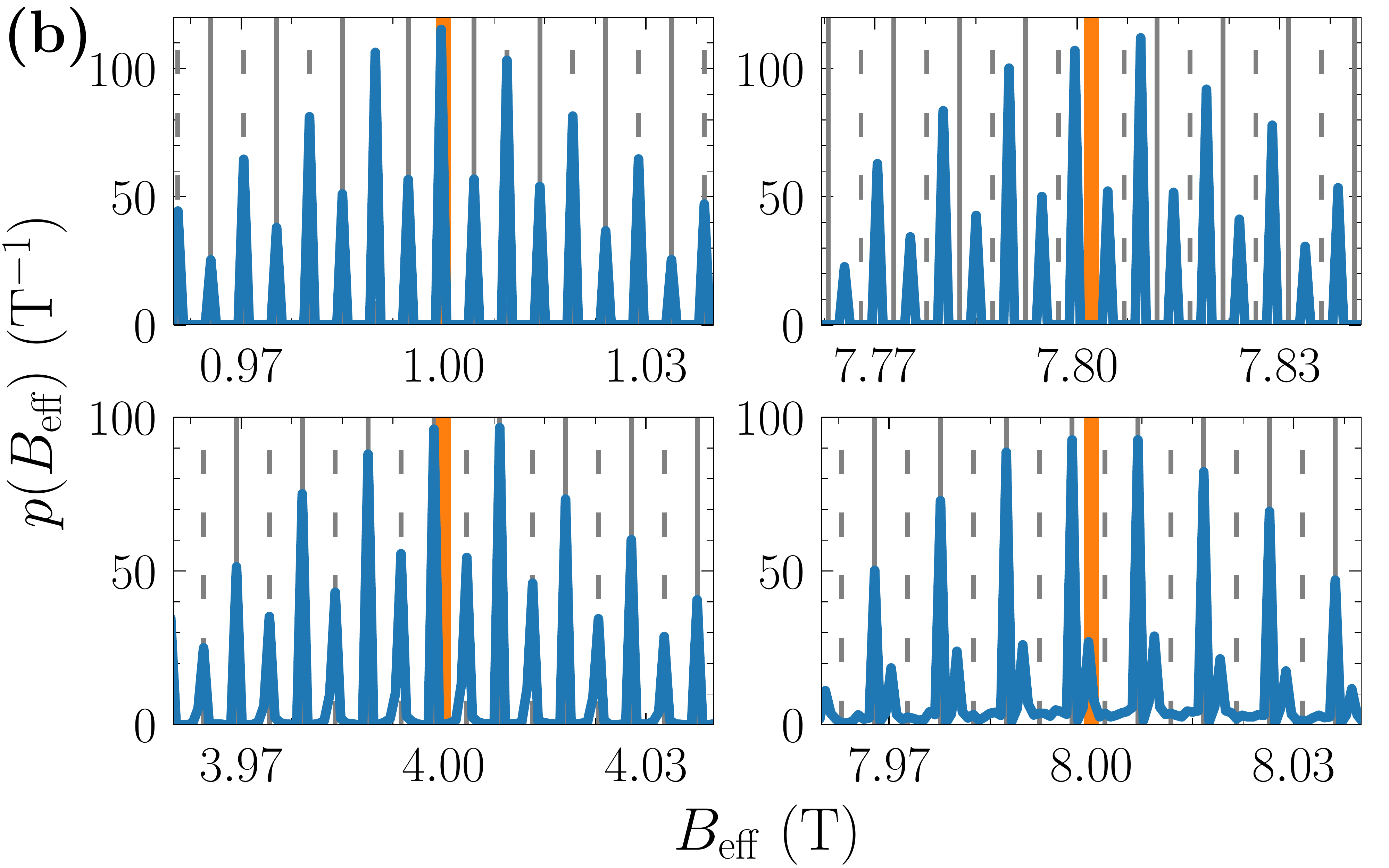} 	\label{fig:TD_Overhauser}}	\\
	\subfloat{\includegraphics[width=\columnwidth]{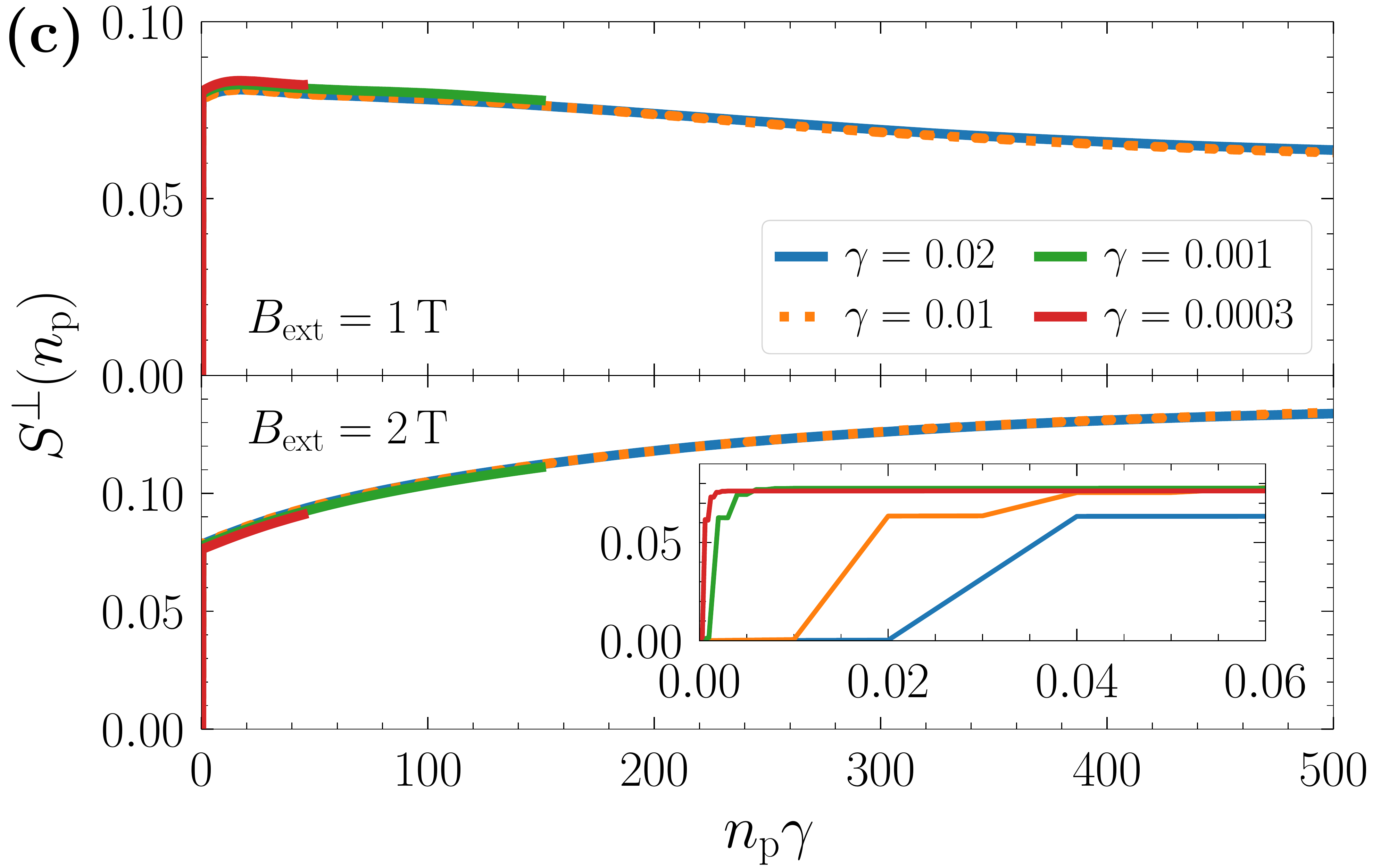} \label{fig:TD_gamma_scaling}}
	\subfloat{\includegraphics[width=\columnwidth]{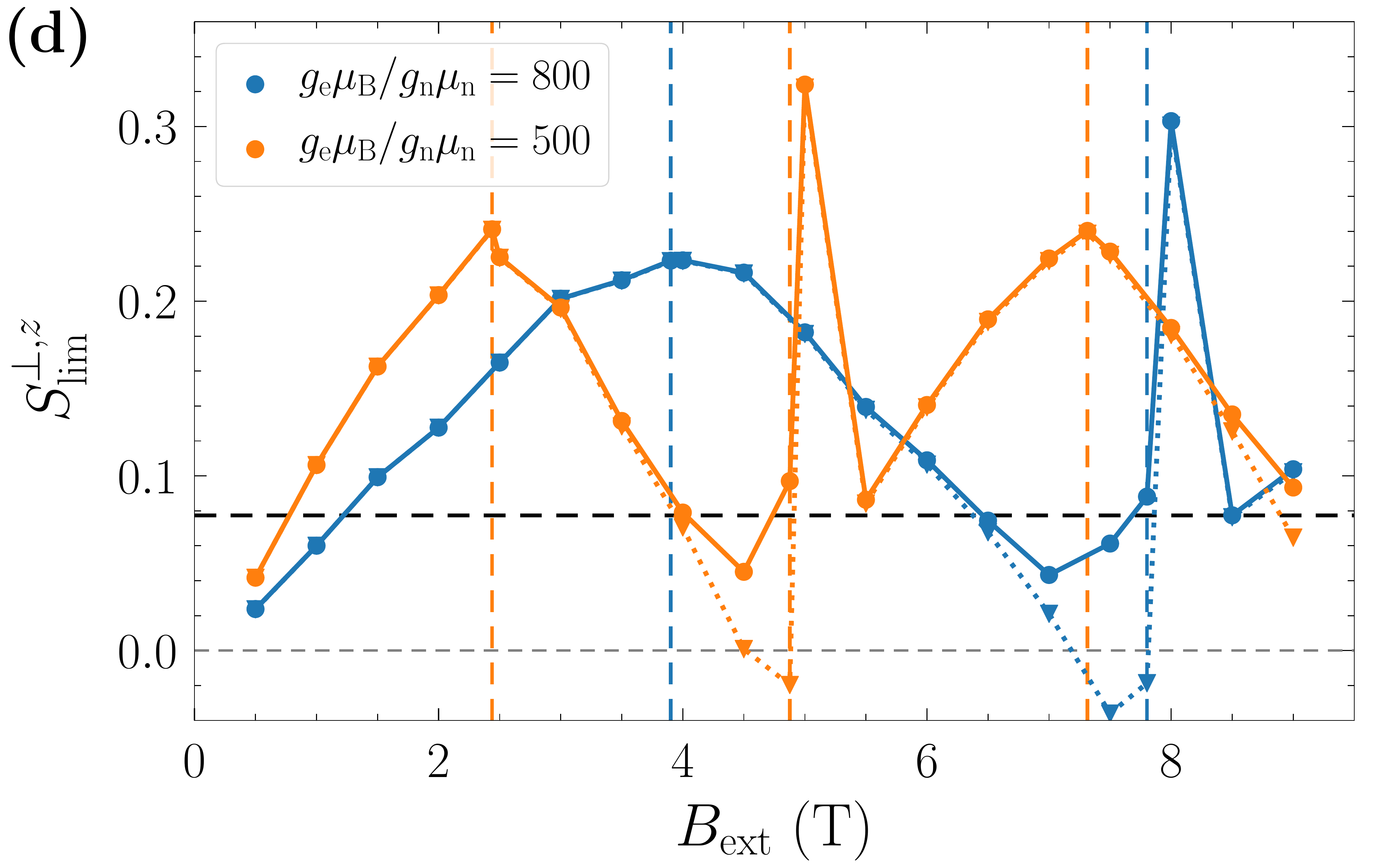} \label{fig:TD_Slim_Bext}}
	\caption{Numerical results for the initial model introduced in Sec.~\ref{sec:initial_model}: 
		(a) Spin dynamics (blue line with fast oscillations) between consecutive pulses after the first pulse, after a few pulses (SML regime), and after many pulses (NIFF regime) for two external magnetic fields $\Bext$. The orange line represents the envelope $S^\perp(t)$.
		(b) Quasi-stationary probability distributions of the effective magnetic field $p(\Beff)$ for various external magnetic fields $\Bext$ (orange vertical lines) and $\gamma = 0.02$. The gray solid and dashed vertical lines represent the values of $\Beff$ fulfilling the ERC~\eqref{eq:ERC} and ORC~\eqref{eq:ORC}, respectively.
		(c) Revival amplitude as a function of the scaled number of pulses~$\np \gamma$ for various values of the coupling parameter $\gamma$; note the data collapse for long trains of pulses. The inset shows the SML regime $\np \gamma \ll 1$.
		(d) Limiting values $S^\perp_\mathrm{lim}$ (sphere, solid) and $S^z_\mathrm{lim}$ (triangle, dotted) of the revival amplitude as a function of the external magnetic field $\Bext$ for two ratios $\ge \muB / \gn \mun$ and $\gamma = 0.02$. The vertical dashed lines represent the corresponding NRCs~\eqref{eq:NRC}, the horizontal black dashed line indicates the SML steady-state value~$\SSML$.
	}
	\label{fig:initial}
\end{figure*}

Under these conditions, the pulse action can be described by a simple relation between the spin components before ($\S_\mathrm{b}$, $\J_\mathrm{b}$) and after ($\S_\mathrm{a}$, $\J_\mathrm{a}$) the pulse~\cite{yugov09,jasch17}
\begin{subequations}
\begin{align}
	S^z_\mathrm{a} &= \frac{1}{4} + \frac{1}{2} S^z_\mathrm{b}, \label{eq:pulse_z}\\
	S^x_\mathrm{a} &= S^y_\mathrm{a} = 0, \\
	J^z_\mathrm{a} &= S^z_\mathrm{b} - S^z_\mathrm{a}, \label{eq:pulse_Jz}\\
	J^x_\mathrm{a} &= J^y_\mathrm{a} = 0 \label{eq:pulse_Jxy},
\end{align}
\label{eq:pulse}%
\end{subequations}
where $J^x$ and $J^y$ are the transverse components of the trion pseudospin vector $\J$. 
The relevant trion spin states can be characterized by an effective pseudospin ${J^z := (T_+ - T_-)/2}$, with $T_\mathrm{\pm}$ being the number of trions with spin projection $\pm 3/2$ (due to the unpaired heavy hole spin) onto the $z$~axis~\cite{glazo12b}.
The transverse components have no relevance in this section as we neglect possible trion spin dynamics here, but they become important later in Sec.~\ref{sec:EM_II}.

Note that for longer pulse durations, the pulse efficiency is reduced especially for very large magnetic fields~\cite{spatzek11,klein18}, but this is beyond the scope of the present work.

\subsection{Results}

We solve the coupled equations of motion~\eqref{eq:eom_S}, \eqref{eq:eom_Jz}, and~\eqref{eq:Qk} for the spin dynamics numerically for $M$ random initial fields $\{ \Q_k \}$ while applying the pulse relation~\eqref{eq:pulse} every $\TR = 13.2$~ns for $\np$~pulses. 
The actual dynamics of the localized electron spin is given by the ensemble average over all $M$ independent trajectories.
In our simulations, we typically use $M = 8192$.

Let us briefly review the basic phenomena of spin mode locking~(SML) and nuclei-induced frequency focusing~(NIFF), which can be already discussed qualitatively for the initial model.
Typical time evolutions between consecutive pulses due to these two effects are shown in Fig.~\ref{fig:TD_S_timeevolution}.
The first pulse creates a net spin polarization which precesses around the transverse magnetic field $\Bext$. Due to the normal distribution of the Overhauser field, this polarization dephases on the timescale $\Tnstar = 1$~ns according to~\cite{merku02,stanek14}
\begin{align}
	S^z_\mathrm{deph} (t) \propto \cos(\ge \muB \Bext t) \exp\left[-\left(\frac{t}{\Tnstar}\right)^2\right].
\end{align}
After applying only a few pulses, a revival of this polarization emerges just before the arrival of the subsequent pulse. This effect is called spin mode locking~\cite{greil06b} and it emerges due to a selection of precession modes due to the properties of the pulse~\eqref{eq:pulse}. 
Qualitatively speaking, the modes corresponding to an integer number of Larmor periods between consecutive pulses lead to an enhancement of spin polarization while the spin polarization is destroyed for modes corresponding to a noninteger number. 
The physics behind this behavior is that due to the selection rules, the localized electron spin is optically inactive in the first case and optically active in the second case.

It can be shown analytically that a steady state emerges for this revival amplitude when neglecting the Overhauser field dynamics and also the trion decay ($\ge \muB \Bext \gg 1/\tau_0$)~\cite{beuge16,jasch17,klein18}.
The steady state follows from the condition $S^z(\np \TR) = S^z(\np \TR + \TR)$ in combination with the periodic application of the pulse~\eqref{eq:pulse}.
After averaging over the Overhauser field distribution, the analytical steady state in the SML regime (without NIFF) takes the value~\cite{beuge16,klein18}
\begin{align}
	\SSML \coloneqq \lim_{\np \to \infty} \overline{S^z(\np\TR^-)} = \frac{1}{\sqrt{3}} - \frac{1}{2} \approx 0.07735, \label{eq:SML_steadystate}
\end{align}
which is also identical to the value of the envelope $S^\perp$ [defined in Eq.~\eqref{eq:Sperp}], i.e., the $x$ and $y$ components vanish.
In the following, we will refer to this value as the SML steady-state value.
This steady state is reached after about ten pulses, independent of the external magnetic field strength.

When driving the system by much longer pulse trains (up to minutes), the effect of nuclei-induced frequency focusing comes into play~\cite{greil07a}, with a rate depending strongly on the parameters $\Bext$, $\gamma$, and $\Tnstar$.
This effect leads to a change of the SML steady-state amplitude.
The periodic driving of the localized electron spin is transferred to the nuclear spin bath via the hyperfine interaction such that the initial normal distributed Overhauser field evolves towards a comb like structure; see Fig.~\ref{fig:TD_Overhauser}. The position of the emerging peaks in the probability distribution of the Overhauser field corresponds to the two resonance conditions~\cite{jasch17,klein18}
\begin{subequations}
\begin{align}
	\Omegaeff \TR &= 2 \pi k, \label{eq:ERC} \\
	\Omegaeff \TR &= 2 \pi k + 2 \arctan \left( \Omegaeff \tau_0 \right) \approx (2k + 1) \pi \label{eq:ORC},
\end{align} \label{eq:resonances}%
\end{subequations}
$k \in \mathbb{Z}$, where ${\Omegaeff = | \Bov + \ge \muB \Bext \ex |}$ is the precession frequency in the effective magnetic field $\Beff = \Omegaeff / \ge \muB$ created by the external magnetic field and the Overhauser field and $\tau_0$ is the radiative lifetime of the trion.
We will refer to the first condition \eqref{eq:ERC} as the \emph{even} resonance condition~(ERC) because $2k$ is an even integer.
The approximation in the second condition \eqref{eq:ORC} holds for $\Omegaeff \tau_0 \gg 1$, which is the case in our theoretical considerations.
Hence, we refer to it as the odd resonance condition~(ORC) because $(2k + 1)$ is an odd integer.
Depending on which resonance condition emerges, the revival amplitude is either increased or decreased with respect to the SML steady-state value $\SSML$.
The correct description of this interplay as a function of the magnetic field is the main goal of this work.

One of the key quantities of interest is the revival amplitude
\begin{align}
	S^\perp(\np) \coloneqq \sqrt{[S^y(\np \TR^-)]^2 + [S^z(\np \TR^-)]^2}, \label{eq:Sperp}
\end{align}
where $\TR^-$ indicates that we take the value of the spin polarization immediately before the arrival of the next pulse.
The trion does not contribute to this revival because it is decayed completely at $t = \np \TR^-$ ($\tau_0 \ll \TR$) .
In particular, we are interested in the long-time behavior of the revival amplitude and its corresponding saturation value
\begin{align}
	\Slim \coloneqq \lim_{\np \to \infty} S^\perp(\np) ,
\end{align}
because this is the state reached in the experiments.
Experimentally, the spin polarization can be probed using weak linearly polarized pulses exploiting the spin Faraday effect.
This yields a signal proportional to $S^z - J^z$~\cite{yugov09,glazo12b}.

Figure~\ref{fig:TD_gamma_scaling} shows the build-up of spin polarization for various values of the parameter $\gamma$. 
The revival amplitude increases to the SML steady-state value $\SSML \approx 0.07735$ within the first few pulses, independent of $\gamma$ and of the magnetic field $\Bext$.
NIFF emerges after a long train of pulses. 
However, there is an important qualitative difference between $\Bext = 1$~T and $2$~T: 
For $\Bext = 1$~T, the revival amplitude decreases, while for $\Bext = 2$~T, we find an increase. 
In both cases, a saturation value is reached eventually.
Scaling the number of pulses $\np$ with $\gamma$ leads to an almost perfect data collapse in Fig.~\ref{fig:TD_gamma_scaling} so that the saturation value $\Slim$ is independent of $\gamma$, differences stem from the statistical nature of our approach.
Note that a similar behavior is found in Refs.~\cite{jasch17,scher18}. 

Since $\gamma \approx 2/N_\mathrm{eff}$ can be interpreted as the inverse effective bath size, we are in particular interested in values $\gamma \approx 10^{-5}$.
However, the computational effort is too big for a direct simulation because the typical coupling strength of a nuclear spin is proportional to $\sqrt{\gamma}$, i.e., the rate of frequency focusing in the nuclear spin bath is much smaller for realistic bath sizes. 
In fact, the rate scales only linearly with $\gamma$ as demonstrated in Fig.~\ref{fig:TD_gamma_scaling} because NIFF is only a second order effect when the nuclear Zeeman term in Eq.~\eqref{eq:Qk} is present~\cite{scher18}.
Then, the nuclear spin dynamics induced by the hyperfine coupling acts as a perturbation with its leading effect occurring in second order.
For this reason, we can study the dependence of $\Slim$ on the magnetic field $\Bext$ for $\gamma = 0.02$~($\Neff \approx 100$) in Fig.~\ref{fig:TD_Slim_Bext}, which is also representative for the limit $\gamma \to 0$.
We also include $S_\mathrm{lim}^z$ to identify possible phase shifts between the signal before and after the pulse.

Previous research has established another class of resonance conditions, namely for the nuclear spins~\cite{beuge17,klein18},
\begin{align}
	\gn \mun \Bext \TR = \pi k, \label{eq:NRC}
\end{align}
$k \in \mathbb{Z}$, which plays a crucial role for the magnetic field dependence of the saturated revival amplitude $\Slim$. 
The values of $\Bext$ fulfilling this condition are highlighted in Fig.~\ref{fig:TD_Slim_Bext} as vertical dashed lines for two different ratios $\ge \muB / \gn \mun$. 
This nuclear resonance condition~(NRC) describes the number of half-turn revolutions of the nuclear spins in the external magnetic field $\Bext$ between consecutive pulses. 
Note that the influence of the small Knight field, i.e., the additional field that a nuclear spin sees due to its coupling to the localized electron spin, is neglected. 
It might induce slight deviations from the expected resonance positions~\cite{uhrig19}.

Let us discuss the details of Fig.~\ref{fig:TD_Slim_Bext}.
We observe that the curve for $\ge \muB / \gn \mun = 500$ is essentially a horizontally rescaled version of the curve for $\ge \muB / \gn \mun = 800$.
Maxima are found close to the values of $\Bext$ fulfilling the NRC~\eqref{eq:NRC}. 
The first maximum ($k=1$, half turn) is rather broad while the second maximum ($k=2$, full turn) is quite sharp and slightly shifted to the right from the expected resonance position. 
We identify the emergence of phase shifts while approaching the second maximum since $\Slim$ and $S^z_\mathrm{lim}$ start to deviate from each other. 
Just after the second resonance is reached, this phase shift vanishes again. 
For $\ge \muB / \gn \mun = 500$, a third maximum ($k=3$) appears, which is very similar to the first one and indicates a periodicity for larger values of $\Bext$.
We expect a third maximum also for the ratio $\ge \muB / \gn \mun = 800$, but it is numerically out of reach.
The heights of the broad maxima are very similar. 
The sharp maximum is slightly less pronounced for $\ge \muB / \gn \mun = 800$, but this is most likely due to the finite discretization of the magnetic field $\Bext$.

The finding of a maximum at $3.9$~T for $\ge \muB / \gn \mun = 800$ is the main downside of this model. Previous research has established theoretical models which predict \emph{minima} at the values of $\Bext$ fulfilling the NRC~\cite{scher18,klein18}, which is also in much better agreement with the experimental results around $\Bext = 4$~T~\cite{jasch17,klein18}.

The structure of the dependence $\Slim(\Bext)$ in Fig.~\ref{fig:TD_Slim_Bext} can be understood by studying the corresponding quasistationary probability distributions of the effective magnetic field~$p(\Beff)$.
The term ``quasistationary'' implies that the distribution does not change noticeably anymore even though the Overhauser field is still dynamic.
A nonequilibrium steady state is reached.
The most characteristic examples for ${\ge \muB / \gn \mun = 800}$ are shown in Fig.~\ref{fig:TD_Overhauser}. 
For almost any external magnetic field $\Bext$, highlighted by the orange vertical line in the plots, we find sharp peaks at the even (vertical solid black lines) and odd (vertical dashed black lines) resonance conditions, but with different weights. 
Since the ERC corresponds to full Larmor periods between consecutive pulses, the steady state condition following from the pulse relation \eqref{eq:pulse_z} is $S{^z_\mathrm{b} = S^z_\mathrm{a} = 1/2}$. For the ORC, the steady state is determined by ${S^z_\mathrm{b} = - S^z_\mathrm{a} = -1/6}$.
For this reason, odd resonances contribute with a three times smaller weight when all contributions from the Overhauser field distribution are summed.
Note that for the same reason the ERC dominates the SML regime without NIFF because the Overhauser field is normally distributed so that this distribution by itself does not favor ERC over ORC or vice versa.

Interestingly, we find strong deviations from the even and odd resonances around $\Bext \approx 7.8$~T, which is the value corresponding to $k = 2$ in the NRC~\eqref{eq:NRC}. This finding explains the deviations of the $z$~component from the envelope in Fig.~\ref{fig:TD_Slim_Bext}.
When increasing the magnetic field just slightly to $\Bext = 8$~T, sharp peaks are found at the ERC, but small side peaks remain which do not correspond to the expected resonances.
At $\Bext = 8.5$~T and $9$~T, the behavior is back to normal with sharp peaks at the ERC and broader peaks at the ORC (not shown).

\subsection{Discussion}

The initial model reveals the fascinating coherent spin phenomena SML and NIFF and also a nonmonotonic dependence $\Slim(\Bext)$, which was first published in Refs.~\cite{varwig14,jasch17} and is studied in more detail in Refs.~\cite{scher18,klein18}.
However, the results for the initial model contradict the findings of previous theoretical research~\cite{klein18} where \emph{minima} instead of maxima were found at the NRC, which is also in much better agreement with the experimental results~\cite{jasch17,klein18}.
While there are experimental results that suggest that both even and odd resonances can appear simultaneously in the frequency spectrum of the localized electron spin, this seems to be not the case for every magnetic field~\cite{jasch17}. 
Moreover, the peaks found in the experiments appear to be generally much broader.
One has to keep in mind that the frequency spectrum of the localized electron spin is not completely equal to the probability distribution of the effective magnetic field, especially due to the different contribution of the ERC and ORC to the spin polarization.
In the following sections, we address the issues of the initial model by extending it in three steps.

\section{Extended model I: Pulse~as~a~measurement}
\label{sec:EM_I}

As we are modeling the system and the pulses in a semiclassical picture, it is not clear how to treat the excitation of the localized electron spin by the circularly polarized laser pulse.
In fact, the pulse model~\eqref{eq:pulse} was derived quantum mechanically~\cite{yugov09}, but the relations are only valid for the expectation values of the spins.
One could argue that applying the pulse relation~\eqref{eq:pulse} to the spin polarization \emph{after} calculating the ensemble average, but this approach destroys any correlation otherwise present in a single configuration such that no revival amplitude appears without NIFF, i.e., the mere SML regime is missing.
However, such correlations are preserved in a quantum mechanical approach.

We extend the pulse model~\eqref{eq:pulse} by interpreting its application as a measurement.
This leads to a nondeterministic pulse description in the sense of a truncated Wigner approximation~(TWA)~\cite{polkovnikov10} and reduces the discrepancy to the fully quantum mechanical behavior.
The same principle was applied in Refs.~\cite{scher18,klein18} to a simpler pulse model, which led to a minimum of the magnetic field dependence $\Slim(\Bext)$ around $4$~T as found in the experiments~\cite{jasch17,klein18}.
Alternative nondeterministic pulse descriptions are discussed in Appendix~\ref{app:pulse_alternatives}, but they turn out to be less reliable.

\subsection{Non-deterministic pulse description}

The essence of simulating quantum mechanics via classical equations of motion is the choice of the correct initial conditions. 
Typically, one tries to fulfill the quantum mechanical moments of the corresponding operators, in our case of the spin operators.
We already apply this principle to the Overhauser field by sampling it from the proper normal distribution. 
In this case, the huge number of nuclear spins forming a spin bath provides a valid justification of the approach based on the central limit theorem~\cite{stanek14}.
In contrast, this argument does not hold for the single localized electron spin which is excited by a pump pulse, so any semiclassical treatment is always an approximation.
Nevertheless, we will show that this procedure leads to promising results in our application.

The TWA is the theoretical foundation of the following approach.
In this semiclassical phase space method, the initial conditions are sampled from the appropriate Wigner distribution, which is eventually truncated by taking only leading order quantum corrections in $\hbar$ into account. 
In leading order, quantum fluctuations appear only through the Wigner distribution of the initial conditions, but they do not affect the equations of motion themselves, i.e., they are classical.
Finally, the quantum mechanical time evolution is mimicked by the ensemble average over all classical trajectories.~\cite{polkovnikov10}

The main requirement is that the nondeterministic pulse retains the properties of pulse~\eqref{eq:pulse} in the SML regime.
We consider \emph{each} pulse to act as a quantum mechanical measurement, i.e., we have to account for the uncertainty principle~\cite{scher18}.
To be precise, the pulse needs to fulfill the quantum mechanical property for spin-$1/2$ operators ${\langle (S^\alpha)^2 \rangle = 1/4}$.
Hence, the deterministic pulse model~\eqref{eq:pulse} is extended to a nondeterministic description in which the electron spin $\S_\mathrm{a}$ after the pulse is sampled from normal distributions characterized by
\begin{subequations}
\begin{align}
	\mathrm{E}[S^z_\mathrm{a}] &= \frac{1}{4} + \frac{1}{2} S^z_\mathrm{b}, \\
	\mathrm{E}[S^x_\mathrm{a}] &= \mathrm{E}[S^y_\mathrm{a}] = 0, \\
	\mathrm{Var}[S^\alpha_\mathrm{a}] &= \begin{cases} \frac{1}{4} - \mathrm{E}^2[S_\mathrm{a}^\alpha], \qquad &\text{if } \mathrm{E}^2[S^\alpha_\mathrm{a}] \le \frac{1}{4}, \\ 0, \qquad &\text{else.} \end{cases}
\end{align}
\label{eq:pulse_distribution}%
\end{subequations}
The distribution is solely determined by the value $S^z_\mathrm{b}$, i.e., the distribution is different for every pulse unless a steady state is reached.
We have to set the variance to zero in some cases because we treat the spins as classical vectors, i.e., a spin component can be larger than $1/2$ due to the sampling from a normal distribution. 
Practically, this issue only arises for the $z$~component, but for about $25\%$ of the pulses. 
This alters its effective variance to a certain extent, but it leaves the expectation value, which is responsible for the correct reproduction of the steady state in the SML regime, unchanged.

The validity of this nondeterministic pulse description is established in Appendix~\ref{app:pulse_alternatives}, where several nondeterministic pulse descriptions are benchmarked in the SML regime against the deterministic pulse model~\eqref{eq:pulse} and its quantum mechanical realization~\cite{klein18}.

Mapping the classical spin vector~$\S$ to a well-defined density matrix of a spin $1/2$ with subsequent application of a pulse operator yields the deterministic pulse relations~\eqref{eq:pulse} for the expectation values of the spin components (see Ref.~\cite{jasch17} for a derivation). The nondeterministic pulse properties must be introduced by hand in a semiclassical approach.
For instance, an average spin polarization $\langle S^\alpha \rangle = 0$ would cause no dynamics in the Overhauser field in a semiclassical approach unless quantum mechanical fluctuations of the localized electron spin are modeled by an appropriate distribution.
This is in line with semiclassical treatments based on truncated Wigner approaches~\cite{polkovnikov10,schachenmayer15} where often normal distributions yield convincing results even though it is known that the Wigner distributions reflecting quantum mechanical measurements involve negative probabilities.

\subsection{Results}

As pointed out above and studied in detail in Appendix~\ref{app:pulse_alternatives}, the nondeterministic pulse description~\eqref{eq:pulse_distribution} does not alter the behavior in the SML regime besides adding additional fluctuations to the spin polarization. 
In the following, we study the interplay of SML and NIFF when long trains of pulses are applied.

\begin{figure*}[t!]
	\centering
	\subfloat{\includegraphics[width=\columnwidth]{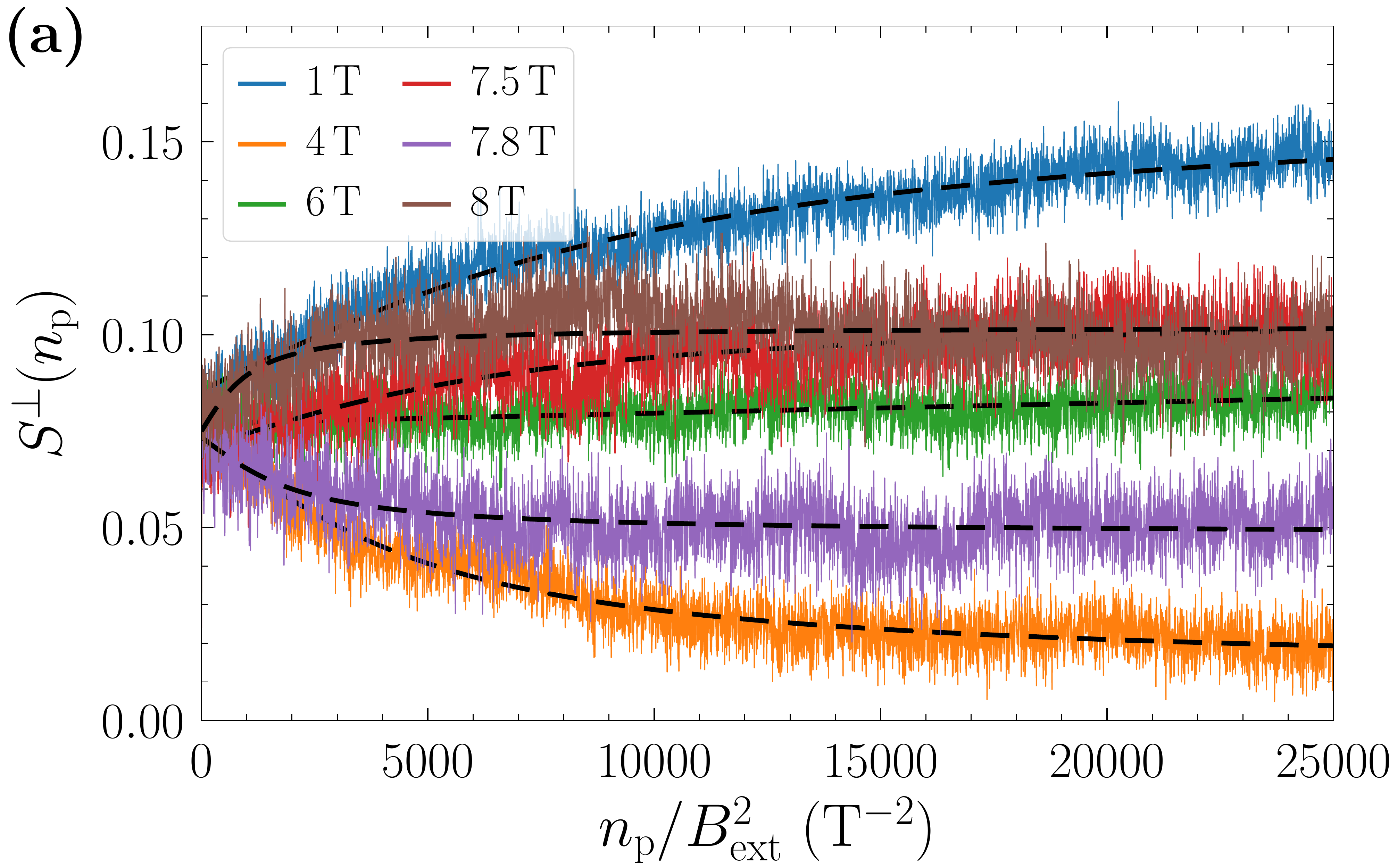} \label{fig:TWA_revival_buildup}}
	\subfloat{\includegraphics[width=\columnwidth]{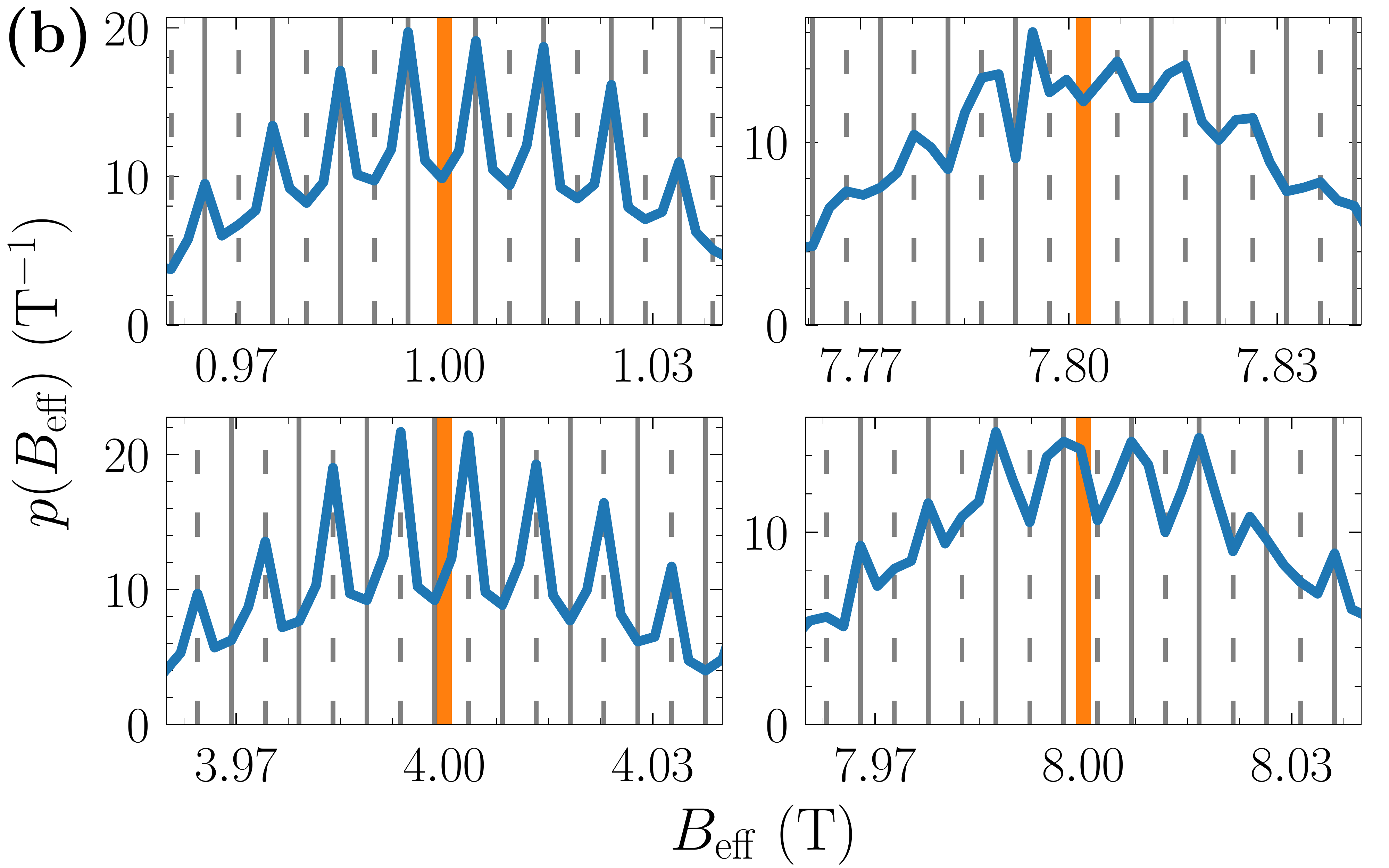} \label{fig:TWA_Overhauser}} \\
	\subfloat{\includegraphics[width=\columnwidth]{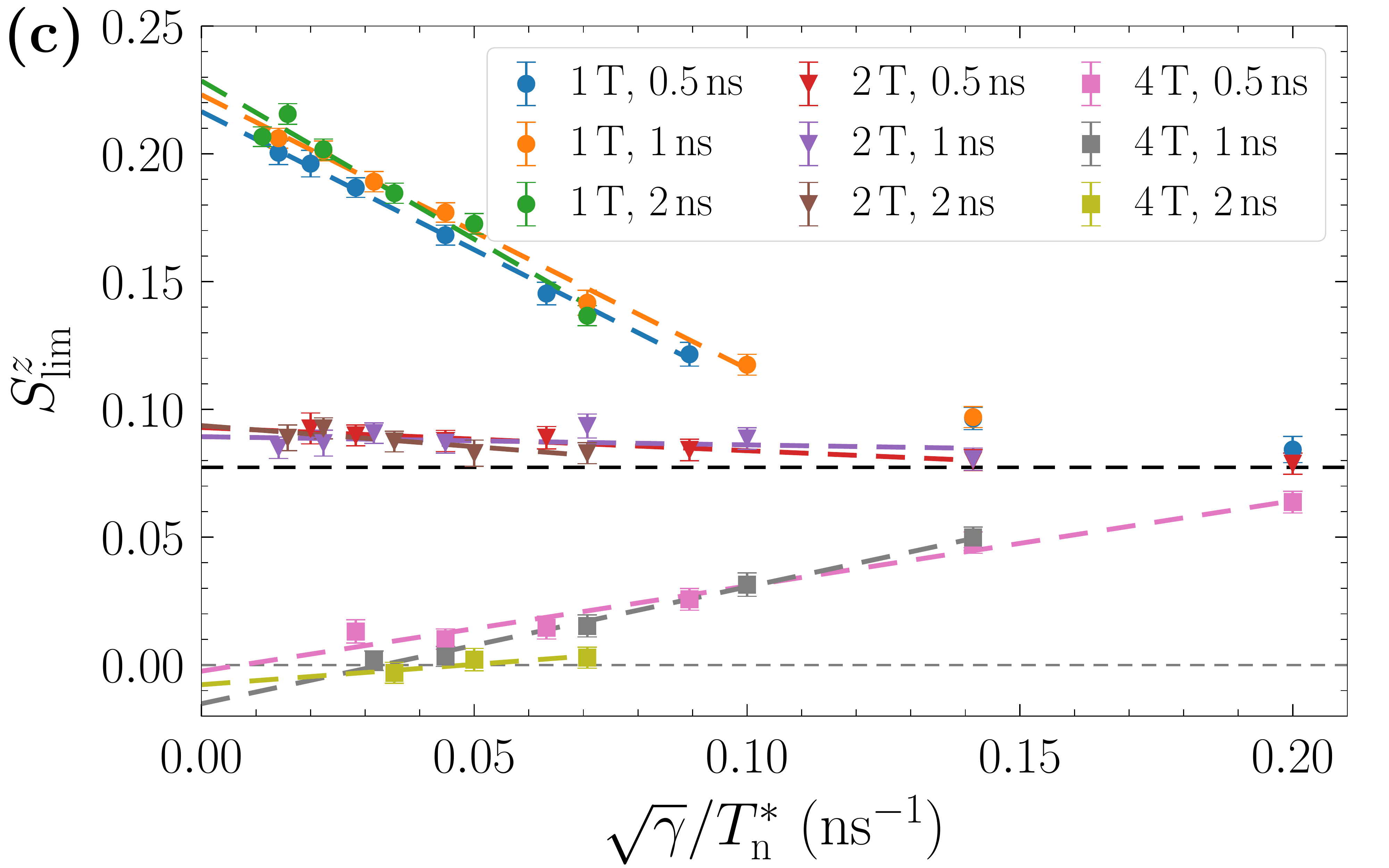} \label{fig:TWA_Slim_gamma}}
	\subfloat{\includegraphics[width=\columnwidth]{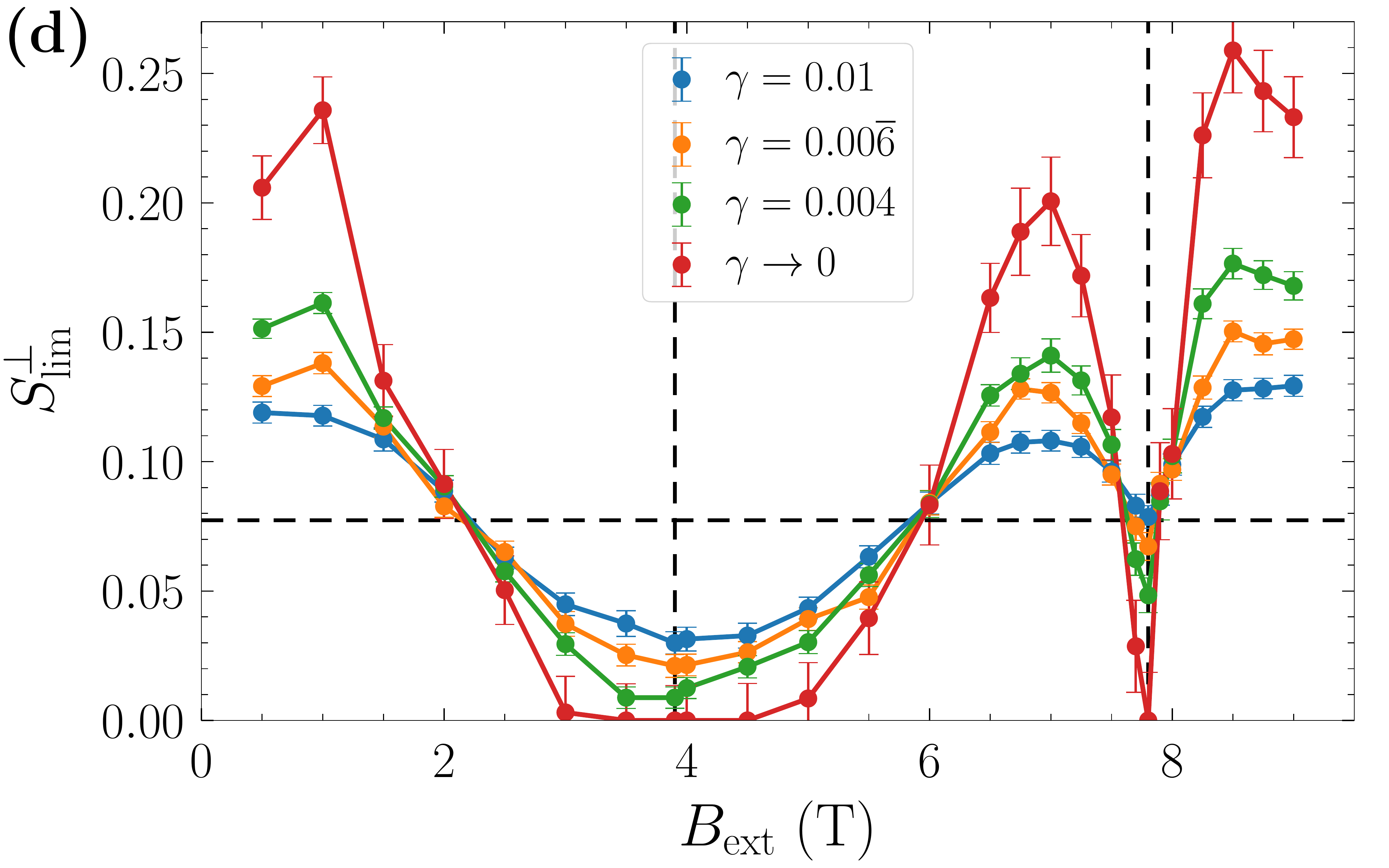} \label{fig:TWA_Slim_Bext}}
	\caption{Numerical results for the extended model~I~(EM~I) introduced in Sec.~\ref{sec:EM_I}: 
		(a) Revival amplitude as a function of the scaled number of pulses $\np / \Bext^2$ for various external magnetic fields $\Bext$ ($\gamma = 0.004$). The black dashed lines represent the corresponding fits~\eqref{eq:Slim_fit}.
		(b) Quasi-stationary probability distributions of the effective magnetic field $p(\Beff)$ for various external magnetic fields $\Bext$ (orange vertical lines) and $\gamma = 0.004$. The gray solid and dashed vertical lines represent the values of $\Beff$ fulfilling the ERC~\eqref{eq:ERC} and ORC~\eqref{eq:ORC}, respectively.
		(c) Saturated revival amplitude $S^\perp_\mathrm{lim}$ as a function of the typical coupling strength $\sqrt{\gamma}/\Tnstar$ for various combinations of the external magnetic field $\Bext$ (in T) and $\Tnstar$ (in ns). The linear extrapolation $\gamma \to 0$ (dashed lines) yields the limit of an infinite bath. The horizontal dashed line represents the SML steady-state value~$\SSML$. 
		(d) Limiting values $S^\perp_\mathrm{lim}$ of the revival amplitude as a function of the external magnetic field $\Bext$ for various values of the coupling parameter $\gamma$ and extrapolated to $\gamma \to 0$. The vertical dashed lines represent the NRC~\eqref{eq:NRC} for $k = 1$~and~$2$, the horizontal dashed line indicates the SML steady-state value~$\SSML$.
	}
	\label{fig:EM_I}
\end{figure*}

Figure~\ref{fig:TWA_revival_buildup} shows the influence of NIFF on the revival amplitude after the SML regime has been reached for various external magnetic fields~$\Bext$.
All curves reach a saturation value after approximately the same number of scaled pulses $\np/\Bext^2$~\cite{scher18,klein18}, but the saturation value depends significantly on the magnetic field strength.
Due to the additional fluctuations of the data caused by the nondeterministic pulse description, we extract the saturation value $\Slim$ by fitting an appropriate function to the data. 
It turns out that
\begin{align}
	S(\np) = A_\mathrm{NIFF} \frac{2}{\pi} \arctan\left( \frac{\np}{\eta} \right) + B_\mathrm{SML} \label{eq:Slim_fit}
\end{align}
is a suitable fit function, which describes a $1/\np$ convergence towards saturation.
This is different from the exponential saturation fit used in Ref.~\cite{klein18}, which would slightly underestimate (overestimate) the saturation value, e.g., for $\Bext = 1$~T ($\Bext = 4$~T).
Fits of type~\eqref{eq:Slim_fit} are included in Fig.~\ref{fig:TWA_revival_buildup} as black dashed lines.
Finally, the saturation value for $\np \to \infty$ is given by ${\Slim = B_\mathrm{SML} + \mathrm{sgn}(\eta) A_\mathrm{NIFF}}$.
Since the fit error turns out to be fairly small, we use the mean-squared error of the last $10\%$ data points as error estimate.
In the rare cases where the fit does not work appropriately, e.g., because almost no NIFF emerges for a given parameter set, we simply interpret the average of the last $10\%$ data points as the saturated revival amplitude.

The quasistationary distributions of the effective magnetic field shown in Fig.~\ref{fig:TWA_Overhauser} reveal much broader peaks than the distributions in Fig.~\ref{fig:TD_Overhauser} for the initial model where the deterministic pulse~\eqref{eq:pulse} is applied.
They are located at the values of $\Beff$ corresponding either to the ERC \emph{or} to the ORC, i.e., only one class of resonances emerges, not both simultaneously as in Fig.~\ref{fig:TD_Overhauser}.

In the extended model, the parameter $\gamma$, which is proportional to the inverse effective bath size, plays an important role.
While the number of pulses required to reach saturation still increases linearly with $\gamma$, the saturation value $\Slim$ changes. 
A similar behavior is found in Ref.~\cite{klein18} for a different, but also nondeterministic pulse when increasing the effective bath size.
It turns out that the typical hyperfine coupling strength, which is proportional to $\sqrt{\gamma}/\Tnstar$, determines the saturation value.
In Fig.~\ref{fig:TWA_Slim_gamma}, we plot $\Slim$ against the typical coupling strength for various combinations of $\Bext$ and $\Tnstar$ while varying $\gamma$.
Especially for $\Bext = 1$~T and $4$~T, there appears to be a linear dependence which we exploit for an extrapolation $\sqrt{\gamma} \to 0$, i.e., to an infinite effective bath size, which is is the limit of interest for QDs with $10^4 - 10^6$ nuclear spins, i.e., $\sqrt{\gamma} \approx 10^{-3} - 10^{-2}$.
Furthermore, the choice of the dephasing time $\Tnstar$, which is an input parameter from the experiments, appears to be not important as long as it is significantly shorter than the pulse repetition time~$\TR$.
Otherwise, one would approach the regime of resonant spin amplification instead of SML, showing different qualitative physics~\cite{yugov12}.
The data and their extrapolation is not as robust for $\Bext = 2$~T where almost no NIFF emerges, but this uncertainty is represented by the fit error being larger than the error of a single saturation value.

For large values of $\sqrt{\gamma}/\Tnstar$, independent of the magnetic field, almost no NIFF emerges and the linear scaling is not applicable.
Physically, the effective hyperfine couplings $\epsilon_k \propto \sqrt{\gamma}/\Tnstar$ becomes too large in comparison to the nuclear Zeeman term $\gn \mun \Bext$ in Eq.~\eqref{eq:Qk}.
Hence, the linear scaling for $\sqrt{\gamma} \to 0$ is expected to work even for larger ratios $\sqrt{\gamma}/\Tnstar$ when a larger magnetic field is applied.
This behavior is evident in Fig.~\ref{fig:TWA_Slim_gamma}.

The minor influence of $\Tnstar$ on the revival amplitude $\Slim$ in the limit $\gamma \to 0$ is very beneficial because the number of pulses required to reach the saturated revival amplitude scales approximately with $(\Tnstar)^3$. 
Thus, we can stick to our initial choice $\Tnstar = 1$~ns without worrying about a strong influence of this parameter which can be also larger for some samples of QDs, e.g., $\Tnstar \approx 4$~ns~\cite{fischer18}. 
Simulations for such a large value are out of reach due to the required computational effort.

We put the new insight to use in Fig.~\ref{fig:TWA_Slim_Bext} where we plot the saturated revival amplitude $\Slim$ as a function of $\Bext$ for decreasing values of $\gamma$. 
Furthermore, we extrapolate the saturation value to an infinite bath size ($\gamma \to 0$) and compare the results to the SML regime.
During this extrapolation process, we enforce the physical lower bound $\Slim \geq 0$.
This is realized by setting $\Slim = 0$ if the extrapolation yields a negative value, but we check that the actual extrapolation value and its error are in agreement with the bound.
We find two \emph{minima} at the values of $\Bext$ which fulfill the NRC~\eqref{eq:NRC}, with the second one being much narrower. 
The minima and maxima become more pronounced for smaller values of $\gamma$, and the minima for $\gamma \to 0$ correspond to $\Slim \approx 0$.
At around $\Bext = 2$~T and $6$~T, almost no NIFF emerges for any choice of $\gamma$.
Between these two values, we find a destructive interplay of SML and NIFF, leading to a decrease of the revival amplitude, and we also find this behavior in a narrow interval around $\Bext = 7.8$~T. 
For the other values of $\Bext$, NIFF increases the revival amplitude, i.e., it acts constructively by enhancing the revival amplitude already present after a few pulses in the mere SML regime.

\subsection{Discussion}

We mimic the quantum mechanical behavior of the system by interpreting each pump pulse as a measurement, leading to a nondeterministic pulse description based on the TWA.
As a result, we find the expected minima in the magnetic field dependence of $\Slim$, similar to the experimental and theoretical results of \citet{klein18}, and in contrast to the initial model studied in the previous section.
Overall, our results are qualitatively very similar to their quantum mechanical approach.
Differences could stem from the considered bath sizes since the quantum mechanical approach is limited to only $N=6$ nuclear spins.

Summarizing, the quasistationary distribution of the effective magnetic field shows much broader peaks than previously in Sec.~\ref{sec:initial_model} for the initial model.
This is in much better agreement with what is reported in experimental studies~\cite{jasch17,evers18}.
Only a single class of resonances emerges, corresponding to either an integer~(ERC) \emph{or} a half-integer~(ORC) number of Larmor periods between consecutive pulses.
Depending on the magnetic field strength, either the ERC or the ORC is fulfilled, which in turn leads to an increase or a decrease of the revival amplitude relative to the SML regime, respectively.
Note that in principle, the emergence of the ORC could also lead to an increase of the revival amplitude relative to the SML steady-state value $\SSML$ if the frequency focusing of the nuclei is strong enough, but we do not observe this behavior in our simulations.
The minima and maxima of the magnetic field dependence become more pronounced for larger bath sizes, the dephasing time $\Tnstar$ only has a minor influence in the limit of an infinite bath size.

\section{Extended model II: Trion~pseudospin~dynamics}
\label{sec:EM_II}

Up to this point, we treated the trion only on the level of an intermediate state which eventually decays on the timescale $\tau_0 = 400$~ps [see Eq.~\eqref{eq:eom}]. 
We neglected the dynamics of its pseudospin $\J$, which is very similar to the dynamics of the ground state of the localized electron spin $\S$ described by the equation of motion~\eqref{eq:eom_S}. 
In the recent theoretical studies of Refs.~\cite{jasch17,scher18,klein18}, this dynamics was not considered either.
But the description of its dynamics, especially the coupling to the external magnetic field, is crucial for the correct description of the time evolution between consecutive pulses.
In the context of spin inertia and polarization recovery measurements, where rather small magnetic fields up to a few $100$~mT are applied in Faraday geometry, the detailed description of the trion pseudospin dynamics is absolutely mandatory~\cite{zhukov18,smirnov18,scher19}. 
In the following, we will show that its inclusion will also alter the behavior of NIFF as a function of the magnetic field, and importantly, we find evidence for dynamic nuclear polarization~(DNP), i.e., the emergence of a nonzero average polarization of the nuclear spin bath, in the model.

\subsection{Equations of motion of the trion pseudospin}

The optical transition induced by a $\sigma^-$ pump pulse leads to the excitation of a singlet $X^-$ trion.
In this case, the trion consists of two electrons in a spin singlet state and a heavy hole with unpaired spin such that the effective type of charge carrier in the excited state (hole) is opposite to the type in the ground state (electron).
We recall that the relevant trion spin states, resulting from the unpaired heavy hole spin, are characterized by an effective pseudospin $J^z := (T_+ - T_-)/2$, with $T_\mathrm{\pm}$ being the number of trions with spin projection $\pm 3/2$ onto the $z$~axis~\cite{glazo12b}.
The dynamics of the corresponding trion pseudospin vector $\J$ is induced by the effective magnetic field, but we need to consider that the hyperfine interaction is much weaker and anisotropic for hole spins because it is caused by the dipole-dipole interaction~\cite{fischer08,testelin09,urba13}.
Then, this interaction can be described by the anisotropic hyperfine Hamiltonian~\cite{hackmann14}
\begin{align}
	\hat{\mathcal{H}}_\mathrm{HF,anisotropic} = \sum_{k=1}^N \left[ \chi A_k \hat{J}^z \hat{I}_k^z + \frac{\chi}{\lambda} A_k ( \hat{J}^x \hat{I}_k^x + \hat{J}^y \hat{I}_k^y) \right].
\end{align}
The resulting classical equation of motion for the trion pseudospin has the form
\begin{align}
	\ddt \J =  \chi \left( B^z_\mathrm{Ov} \ez + \frac{1}{\lambda} \Bov^\perp \right) &\times \J \nonumber \\
	+ \,\, \gh \muB \Bext \ex &\times \J - \frac{1}{\tau_0} \J, 
	\label{eq:eom_fT}%
\end{align}
with $\Bov^\perp = B_\mathrm{Ov}^x \ex + B_\mathrm{Ov}^y \ey$.
The factor $\chi$ describes how much weaker the hyperfine interaction is for hole spins than for electron spins. 
Typically, it is about five to ten times weaker~\cite{urba13,glase16,zhukov18} and thus, we use $\chi = 0.2$ in our calculations.
For the anisotropy factor, $\lambda = 5$ is a typical value~\cite{glase16,zhukov18}.
The $g$ factor of the trion pseudospin (heavy-hole) ranges from $\gh = 0.05$ to $0.15$, depending on the in-plane orientation of the QD sample~\cite{yugov07,crooker10}.
We focus on $\gh = 0.15$ here and demonstrate in Appendix~\ref{app:gh_comparison} that the results are very similar for $\gh = 0.05$.

The equations of motion for the Overhauser field $\Bov$ dynamics are also extended due to its coupling to the trion pseudospin during the trion lifetime $\tau_0$.
The weaker and anisotropic hyperfine interaction is accounted for by the parameters $\lambda$ and $\chi$, leading to the extended equation of motion for the auxiliary vectors
\begin{align}
	\ddt \Q_k = \epsilon_k \bm S \times \Q_k &+ \chi \epsilon_k \left( J^z \ez + \frac{1}{\lambda} \J^\perp \right) \times \Q_k \nonumber\\
	&+ \gn \muB \Bext \ex \times \Q_k, \label{eq:eom_Q_fT}
\end{align}
$k \in \{1,2,\dots,\Ntr\}$, with $\J^\perp = J^x \ex + J^y \ey$. 
The $\epsilon_k$ and $W_k$ do not change as the hyperfine couplings $A_k$ are still parameterized by the exponential distribution~\eqref{eq:couplings}, the Overhauser field is still given by Eq.~\eqref{eq:Overhauser_SDA_Appendix}.

\begin{figure*}[t!]
	\centering
	\subfloat{\includegraphics[width=\columnwidth]{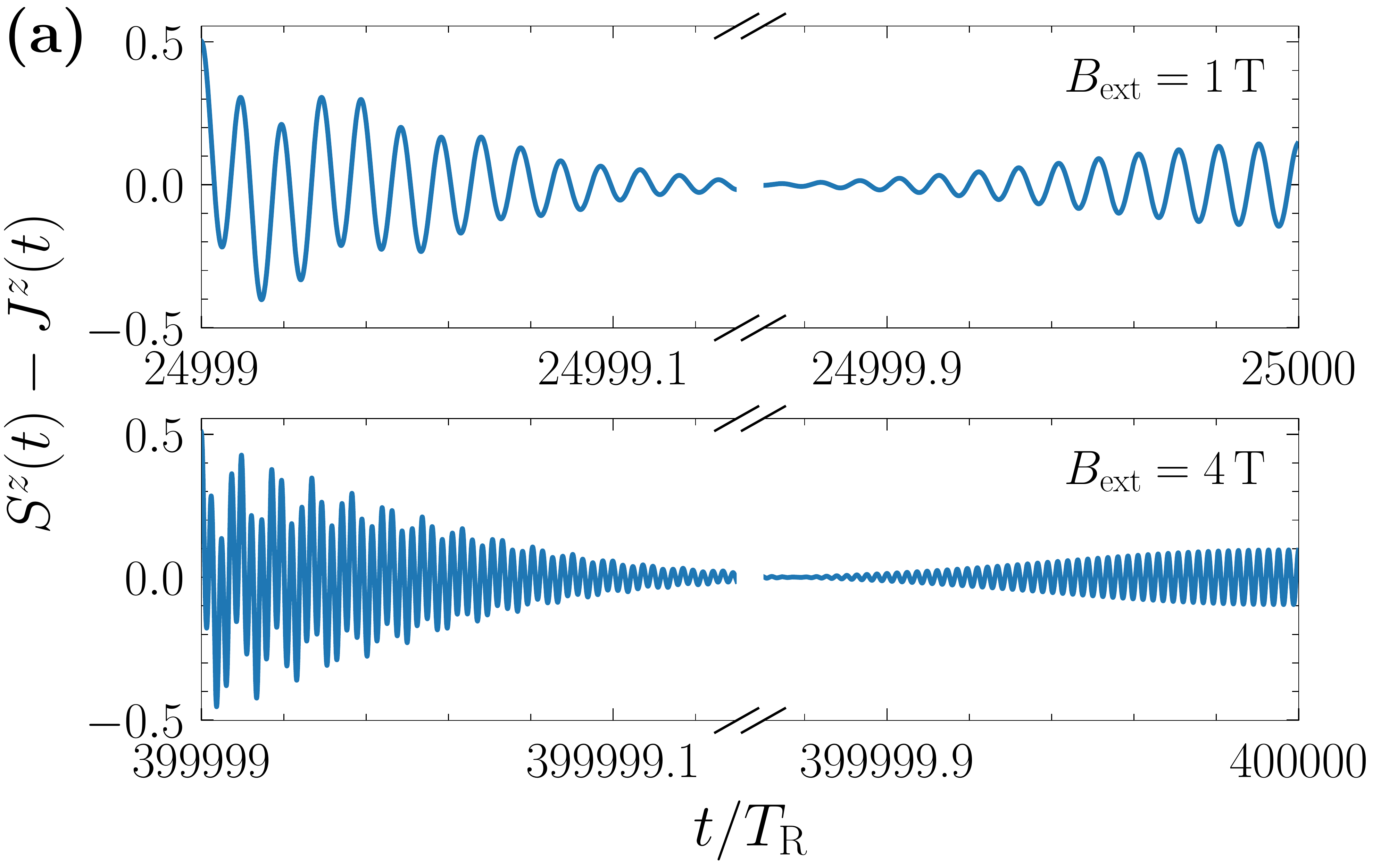} \label{fig:TWAfT_S_timeevolution}}
	\subfloat{\includegraphics[width=\columnwidth]{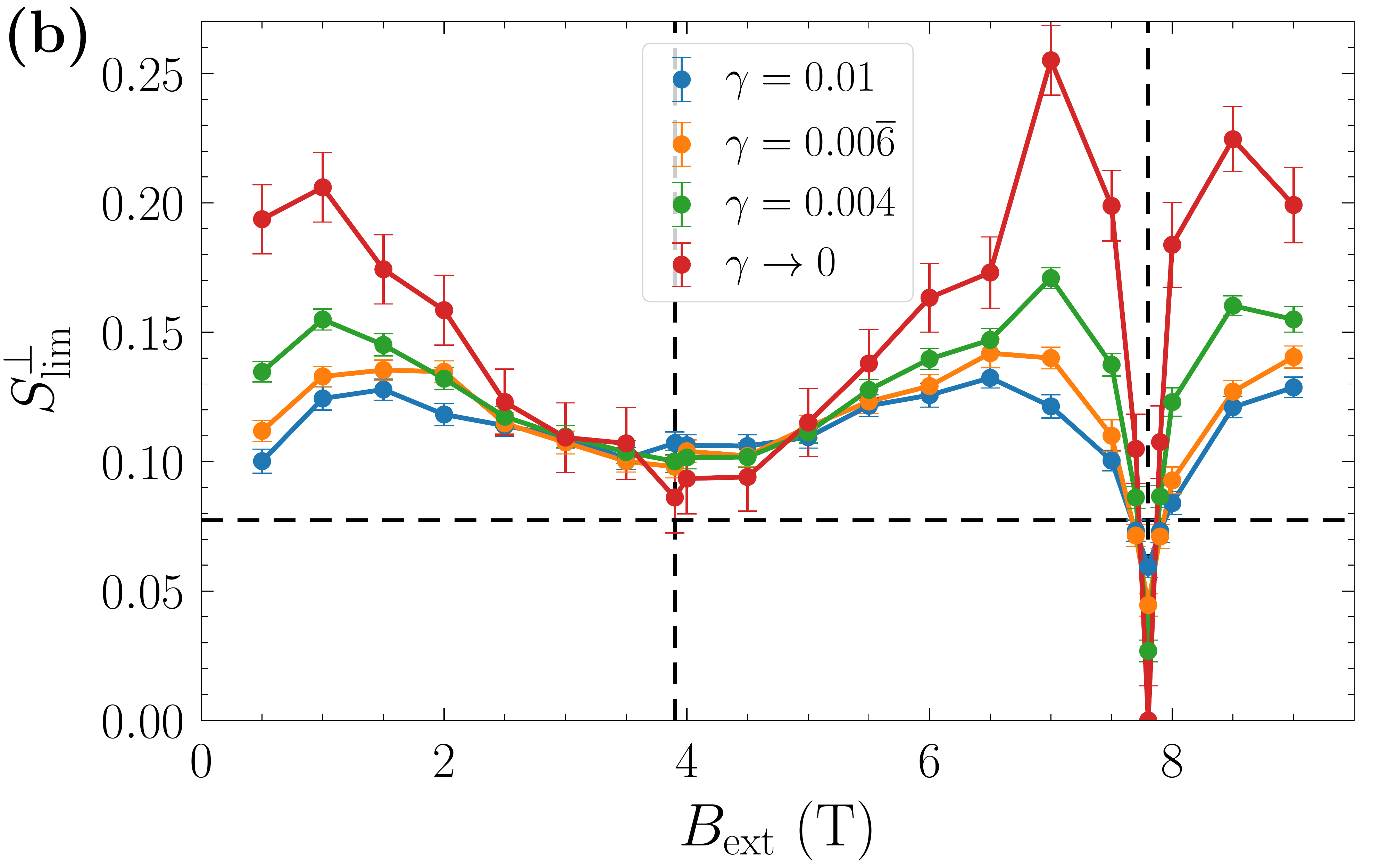} \label{fig:TWAfT_Slim_Bext}} \\
	\subfloat{\includegraphics[width=\columnwidth]{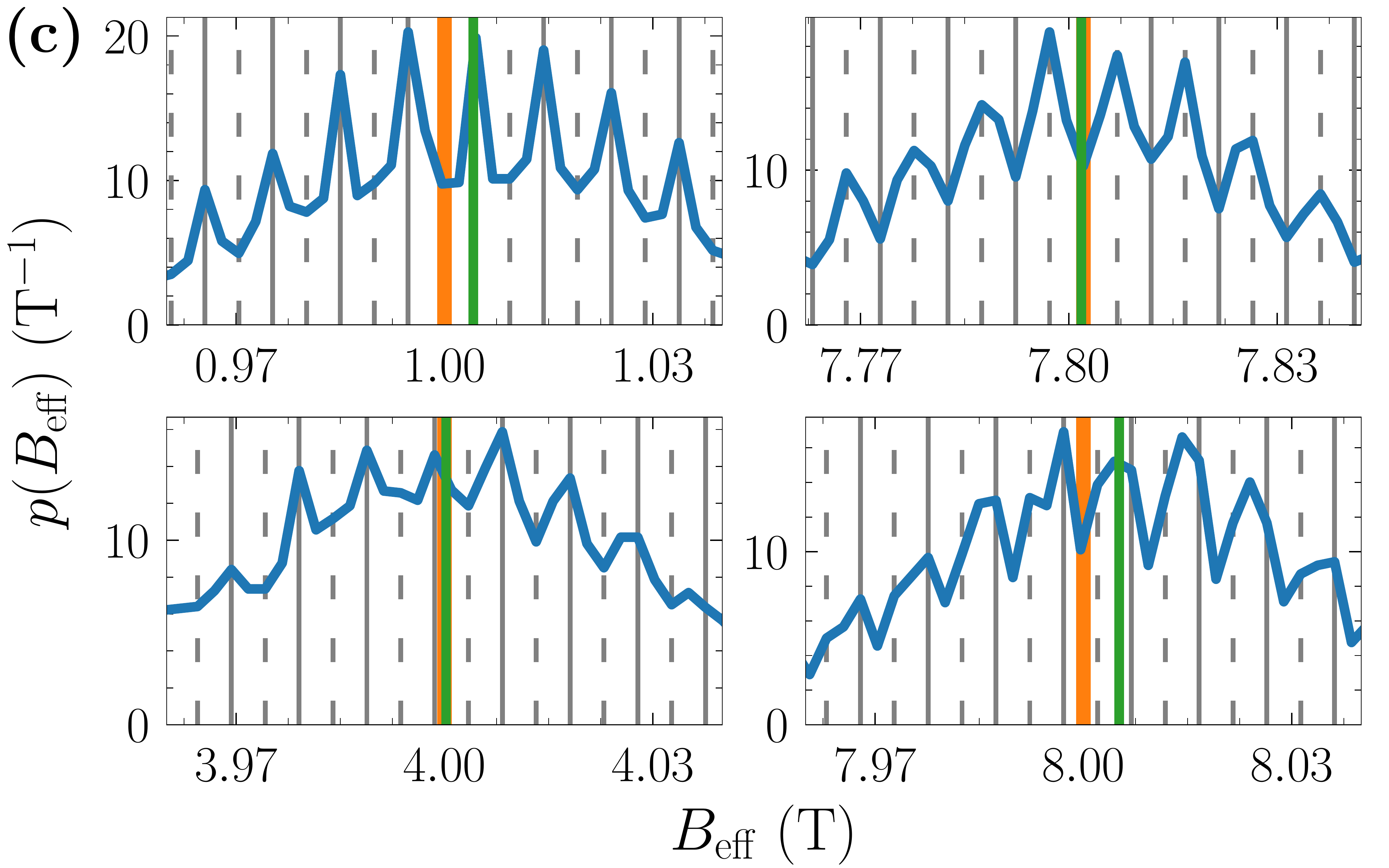} \label{fig:TWAfT_Overhauser}}
	\subfloat{\includegraphics[width=\columnwidth]{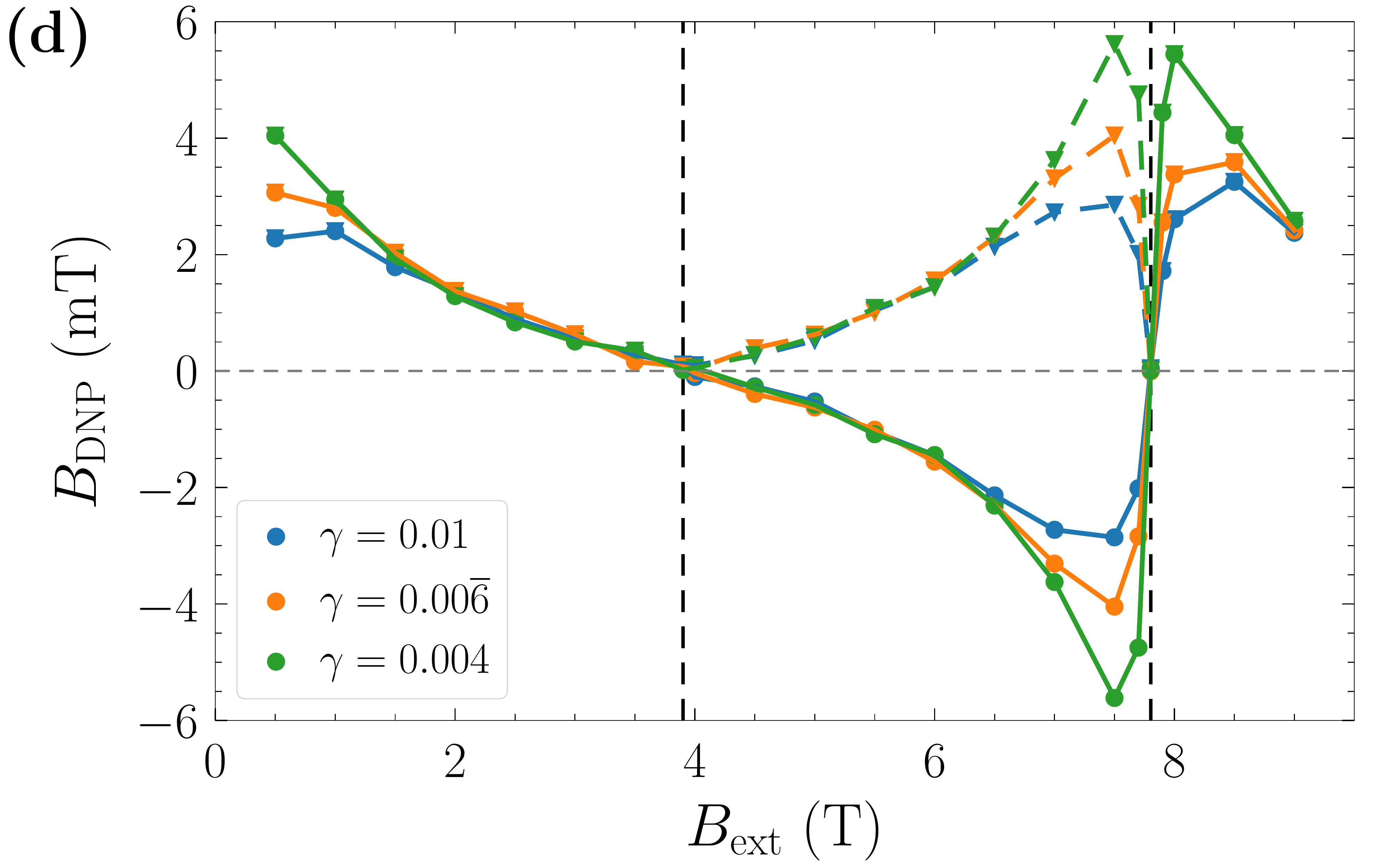} \label{fig:TWAfT_DNP}}
	\caption{Numerical results for the extended model~II~(EM~II) introduced in Sec.~\ref{sec:EM_II}: 
		(a) Spin dynamics between consecutive pulses after a long train of pulses for two external magnetic fields $\Bext$ at $\gamma = 0.004$.	
		(b) Limiting values $S^\perp_\mathrm{lim}$ of the revival amplitude as a function of the external magnetic field $\Bext$ for various values of the coupling parameter $\gamma$ and for the limit $\gamma \to 0$. The vertical dashed lines represent the NRC~\eqref{eq:NRC} for $k = 1$~and~$2$s, the horizontal dashed line indicates the SML steady-state value~$\SSML$.
		(c) Probability distributions of the effective magnetic field $p(\Beff)$ when the revival amplitude is in the saturation regime for various external magnetic fields $\Bext$ (orange vertical lines) and $\gamma = 0.004$. The gray solid and dashed vertical lines represent the values of $\Beff$ fulfilling the ERC~\eqref{eq:ERC} and ORC~\eqref{eq:ORC}, respectively. The green vertical line represents the mean value of the distribution.
		(d) DNP $\Bdnp$~(circles, solid lines) and its absolute value~(triangles, dashed lines) as a function of the external magnetic field $\Bext$ for various values of the coupling parameter $\gamma$ after the saturation of the revival amplitude is reached. The vertical dashed lines represent the corresponding NRCs~\eqref{eq:NRC}.	
	}
	\label{fig:EM_II}
\end{figure*}

\subsection{Results}

The numerical integration of the extended equations of motion~\eqref{eq:eom_fT} and~\eqref{eq:eom_Q_fT} leads only to a negligible increase of computational complexity in comparison to the equations of motion considered in the two previous sections. 
In the following, we will discuss the main changes which emerge in comparison to the results for the EM~I of Sec.~\ref{sec:EM_I}.

\subsubsection{Spin dynamics}

In the pump-probe experiments under consideration, the Faraday rotation or ellipticity is measured by weak linearly polarized pulses. The probed signal is proportional to $S^z - J^z$~\cite{yugov09}, i.e., the spin polarization of the system is measured.
Figure~\ref{fig:TWAfT_S_timeevolution} shows the corresponding time evolution in our model between two pulses in the saturation regime for the two magnetic fields $\Bext = 1$~T and $4$~T. 
The initial dephasing reveals additional beats, stemming from the trion pseudospin which precesses around the external magnetic field with a slightly different Larmor frequency than the electron spin.
The beats decay on the timescale $\tau_0 \ll \TR$ so that they do not appear in the revival before the next pulse.
They are also evident in the experimental results~\cite{greil06a,greil06b,greil07a}, but they typically vanish much quicker there.
This issue is dealt with in Sec.~\ref{sec:EM_III}.

\subsubsection{Nuclei-induced frequency focusing}

The most prominent difference to the EM~I of Sec.~\ref{sec:EM_I} appears when studying the magnetic field dependency of the revival amplitude, which is plotted in Fig.~\ref{fig:TWAfT_Slim_Bext} for various values of the parameter $\gamma$. 
While we still find two minima at values of $\Bext$ fulfilling the NRC~\eqref{eq:NRC}, the first broad minimum hints at the emergence of even resonances instead of the previous odd ones because the value of the revival amplitude is larger instead of smaller compared to $\SSML$.
This is supported by the corresponding quasistationary distribution of the effective magnetic field $p(\Beff)$ for $\Bext = 3.9$~T in Fig.~\ref{fig:TWAfT_Overhauser} where peaks are found at the values of $\Bext$ fulfilling the ERC~\eqref{eq:ERC}.
In this regime, the values of the revival amplitude are larger than the mere SML steady-state value~$\SSML$, which is plotted as a black horizontal dashed line in Fig.~\ref{fig:TWAfT_Slim_Bext}.
The overall degree of NIFF in this regime is small.
This behavior differs from our findings for the EM~I, where the revival amplitude approaches zero and the effect of NIFF becomes more pronounced for small values of $\gamma$.
Furthermore, the second minimum in Fig.~\ref{fig:TWAfT_Slim_Bext} still corresponds to the ORC~\eqref{eq:ORC} in the probability distribution $p(\Beff)$ as shown in Fig.~\ref{fig:TWAfT_Overhauser} for $\Bext = 7.8$~T. 
This minimum is narrower compared to the second minimum found for the EM~I.

The linear extrapolation $\sqrt{\gamma} \to 0$ for the saturation value $\Slim$, which we have established in Fig.~\ref{fig:TWA_Slim_gamma} of Sec.~\ref{sec:EM_I}, is still applicable when including the full trion pseudospin dynamics, and the exact choice of the dephasing time $\Tnstar$ also has only a minor influence on the results.
We apply the extrapolation procedure for the magnetic field dependence $\Slim(\Bext)$ of the revival amplitude in Fig.~\ref{fig:TWAfT_Slim_Bext}.
Overall, the structure becomes more pronounced in the limit $\gamma \to 0$, but the revival amplitudes around the minimum at $\Bext = 3.9$~T are almost independent of $\gamma$.
Moreover, the maxima have a similar height as in Fig.~\ref{fig:TWA_Slim_Bext}.
This means that the degree of NIFF under optimal conditions is very similar.
Under these conditions, the revival amplitude is about three times larger than in the SML regime without NIFF.
Experimentally, a ratio of 3.6 for the revival amplitude with versus without NIFF is found for $\Bext = 2$~T~\cite{evers18}.
For this particular magnetic field, we find a ratio of only 2.
Note that for the EM~I (Fig.~\ref{fig:TWA_Slim_Bext}), this ratio is barely larger than 1.

The influence of the parameter $\chi = 0.2$, which characterizes the hyperfine interaction strength between the trion pseudospin and the Overhauser field, is minor. 
Since the trion only has a lifetime of $400$~ps and the hyperfine interaction is much weaker than for the electron spin, this coupling could be neglected for its smallness.
In our simulations, we find very similar NIFF for $\chi = 0$, i.e., without the coupling of the trion pseudospin to the Overhauser field. 
At best, NIFF is marginally more pronounced when this coupling is neglected because it acts as a small additional perturbation.
However, the deviations between the results are of the order of the estimated error so that no reliable conclusion can be given.

The distribution of the hyperfine couplings is also not essential for the qualitative NIFF behavior.
When we employ a simple box model, i.e., we choose all hyperfine couplings to be equal with $A_k \propto (\sqrt{N} \Tnstar)^{-1}$, the magnetic field dependence of the revival amplitude $\Slim(\Bext)$ shows the same qualitative shape as in Fig.~\ref{fig:TWAfT_Slim_Bext}, with a slightly more pronounced NIFF.
Since all nuclear spins precess with the same frequency in this simplified model, the nuclear spins cannot change their mutual angles. 
We will see, however, that the alignment of nuclear spins is one of the two essential mechanisms leading to dynamic nuclear polarization.

\subsubsection{Dynamic nuclear polarization}

In the probability distributions of the effective magnetic field $p(\Beff)$ in Fig.~\ref{fig:TWAfT_Overhauser}, one can discern a small shift of the distribution to the right for $\Bext = 1$~T and $8$~T.
We highlight this shift by including the mean value $\overline{\Beff}$ of the distribution as a green vertical line in the plots.
The applied magnetic field $\Bext$ is highlighted in orange. 
Remember that at the beginning of each simulation, $\overline{\Beff} \approx \Bext$ holds, with small deviations stemming only from the statistical nature of our approach.

The shift results from dynamic nuclear polarization~(DNP) in the Overhauser field, i.e., the nuclear spins align along the axis of the external magnetic field $\Bext \ex$ to a certain extent.
First, there is the possibility of an internal alignment of the nuclear spins in each QD.
Second, the Overhauser fields of all QDs in the ensemble could also align. 
The second external process can be result of the first one, but the internal alignment of nuclear spins is not possible when a simple box model is used for the hyperfine couplings, i.e., when they are all set equal.

In order to analyze this phenomenon in more detail, we define the DNP as
\begin{align}
	\Bdnp(\np) \coloneqq \frac{\overline{B_\mathrm{Ov}^x(\np \TR)} - \overline{B_\mathrm{Ov}^x(\np \TR = 0)}}{\ge \muB}
	\label{eq:DNP_shift}
\end{align}%
and study it as a function of the magnetic field for several values of $\gamma$ in Fig.~\ref{fig:TWAfT_DNP}. 
The number of pulses $\np$ is chosen such that $\Slim$ is approximately in saturation.
The dots and solid lines represent $\Bdnp$, the triangles and dashed lines its absolute value.
The dependence has a very similar shape to the one of $S^\perp_\mathrm{lim}$ in Fig.~\ref{fig:TWAfT_Slim_Bext} when studying the absolute value of the shifts, and we find no DNP at the magnetic fields fulfilling the NRC~\eqref{eq:NRC}.
This suggests that the underlying mechanisms have a similar origin in the equations of motion.

The difference between the DNP for different values of $\gamma$ is minor for most magnetic fields.
Note that the DNP $\Bdnp$ plotted in Fig.~\ref{fig:TWAfT_DNP} do not represent the stationary values even though the values of $\Slim$ are approximately saturated because the DNP approaches its limit much slower than the revival amplitude.
This leads to slightly different results for the different values of $\gamma$. 
It appears that for the magnetic fields for which the DNP $\Bdnp$ is most prominent, smaller values of $\gamma$ correspond to a stronger DNP.
Unfortunately, it is not possible to reach the stationary DNP regime for large magnetic fields due to its slow convergence, but we analyze the precise DNP behavior for small magnetic fields in the following.

\begin{figure}[b!]
	\centering
	\includegraphics[width=\columnwidth]{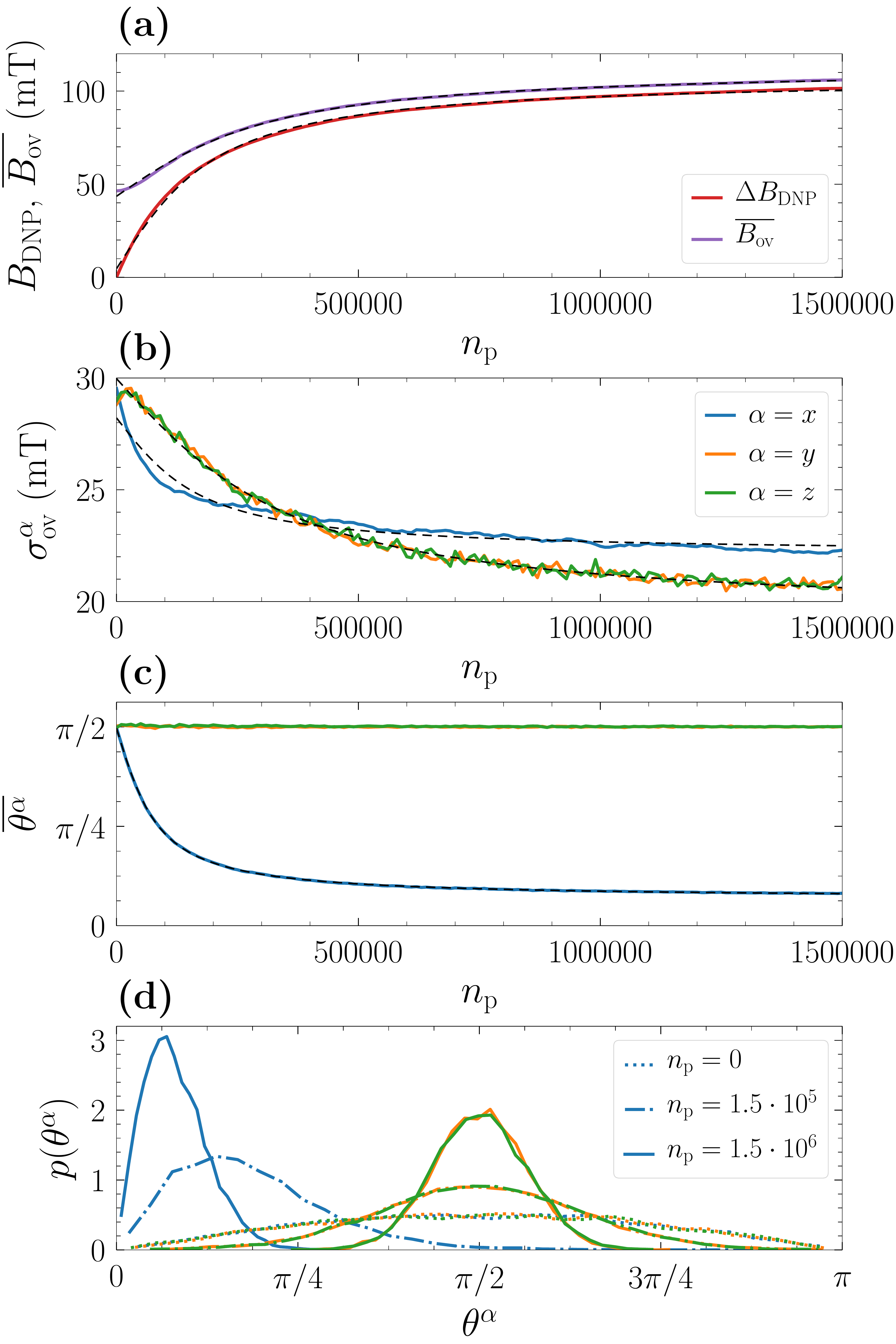}
	\caption{Analysis of the DNP behavior in the extended model~II~(EM~II) for $\Bext = 0.5$~T and $\gamma = 0.004$ due to periodic driving with pulses. All fits (black dashed lines) are of the type~\eqref{eq:arctan}.
	(a) Build-up of the DNP $\Bdnp$ (blue line), defined by Eq.~\eqref{eq:DNP_shift}. The orange line shows the simultaneous increase of the average Overhauser field length $B_\mathrm{Ov}$.
	(b) Decrease of the standard deviation of the Overhauser field components $B_\mathrm{Ov}^\alpha$, $\alpha \in \{x,y,z\}$.
	(c) Average angles $\overline{\theta^\alpha}$ between the Overhauser field and the unit vectors $\bm e_\alpha$. 
	(d) Probability distributions of the angles $p(\theta^\alpha)$ after different numbers of pulses.
	}	\label{fig:TWAfT_DNP_combined}
\end{figure}

In Fig.~\ref{fig:TWAfT_DNP_combined}, we analyze the saturation behavior of DNP and the internal mechanism leading to its emergence.
Figure~\ref{fig:TWAfT_DNP_combined}a shows the build-up of the DNP $\Bdnp$ due to periodic driving with pulses for $\Bext = 0.5$~T and $\gamma=0.004$, eventually reaching a steady state of about $100$~mT after more than $10^6$ pulses. 
This DNP is fairly large in comparison to the initial standard deviation of the Overhauser field components of about $29$~mT.
Note that the revival amplitude $\Slim$ is already saturated after about $5000$ pulses for this set of parameters.
A longer pulse train which is two orders of magnitude longer is required to reach the saturation regime for DNP.

The DNP build-up and saturation as a function of the number of pulses can be described via
\begin{align}
	f_\mathrm{DNP} (\np) = A_\mathrm{DNP} \frac{2}{\pi} \arctan\left( \frac{\np}{\eta}\right) + B_\mathrm{DNP},
	\label{eq:arctan}
\end{align}
where $A_\mathrm{DNP}$, $B_\mathrm{DNP}$, and $\eta$ are fit parameters. 
This function shows a $1/\np$ convergence towards saturation.

In parallel, the average length of the Overhauser field 
\begin{align}
	\overline{B_\mathrm{Ov}} \coloneqq \frac{\overline{|\Bov|}}{\ge \muB}
\end{align}
increases due to an alignment of the individual nuclear spins.
Its dependence on the number of pulses can be described by the fit~\eqref{eq:arctan} as well.
However, this lengthening alone does not explain the emergence of DNP completely; see below.

Figure~\ref{fig:TWAfT_DNP_combined}c shows the average angles 
\begin{align}
	\overline{\theta^\alpha} \coloneqq \overline{\arccos\left(\frac{B^\alpha_\mathrm{Ov}}{|\Bov|}\right)},
\end{align}
$\alpha \in \{x,y,z\}$, between the Overhauser field components $B^\alpha_\mathrm{Ov}$ and the unit vectors $\bm e_\alpha$ as a function of the number of pulses.
The average of the initial angle is given by $\pi/2$ for all components $\alpha$ since the components $B^\alpha_\mathrm{Ov}$ are sampled from a simple normal distribution. 
Driving the system with periodic pulses does not influence the average angles $\overline{\theta^y}$ and $\overline{\theta^z}$, but the average angle $\overline{\theta^x}$ is reduced to about $\pi/12$, implying that the Overhauser field aligns along the $x$ axis, i.e., the axis of the external magnetic field. 
The dependence of $\overline{\theta^x}$ on the number of pulses can be described again by the function~\eqref{eq:arctan}.
Note that $\overline{\theta^x}$ is not reduced to zero because of the finite components $B_\mathrm{Ov}^y$ and $B_\mathrm{Ov}^z$.
These components still follow a normal distribution but with a reduced standard deviation.

The corresponding probability distributions of the angles $\theta^\alpha$ after different number of pulses are plotted in Fig.~\ref{fig:TWAfT_DNP_combined}d.
Initially, all components follow the same distribution with a maximum at $\pi/2$. 
Due to the continuous driving with pulses, the distributions of all components become narrower, i.e., more focused around a certain angle.
The angles $\theta^y$ and $\theta^z$ remain centered around $\pi/2$, but the angle $\theta^z$ becomes significantly smaller, which is also shown in Fig.~\ref{fig:TWAfT_DNP_combined}c.

An important consequence of DNP is a narrowing of the Overhauser field distribution which is demonstrated in Fig.~\ref{fig:TWAfT_DNP_combined}b.
The standard deviations of the Overhauser field components 
\begin{align}
	\sigma^\alpha_\mathrm{Ov} \coloneqq \frac{\sqrt{\mathrm{Var}[B_\mathrm{Ov}^\alpha]}}{\ge \muB},
\end{align} 
$\alpha \in \{x,y,z\}$, are reduced from their initial value of about $29$~mT to about $21-22$~mT. 
This process is faster for the $x$ component and the precise saturation value differs slightly from the one for the $y$~and~$z$~components.
Importantly, this narrowing of the Overhauser field distribution implies an increase of the coherence. The dephasing time is inversely proportional to the standard deviation of the Overhauser field components according to the definition~\eqref{eq:Overhauser_variance}.
Note that a fit with the function~\eqref{eq:arctan} works well again, allowing for an extrapolation $\np \to \infty$.

\begin{figure}[t!]
	\centering
	\includegraphics[width=\columnwidth]{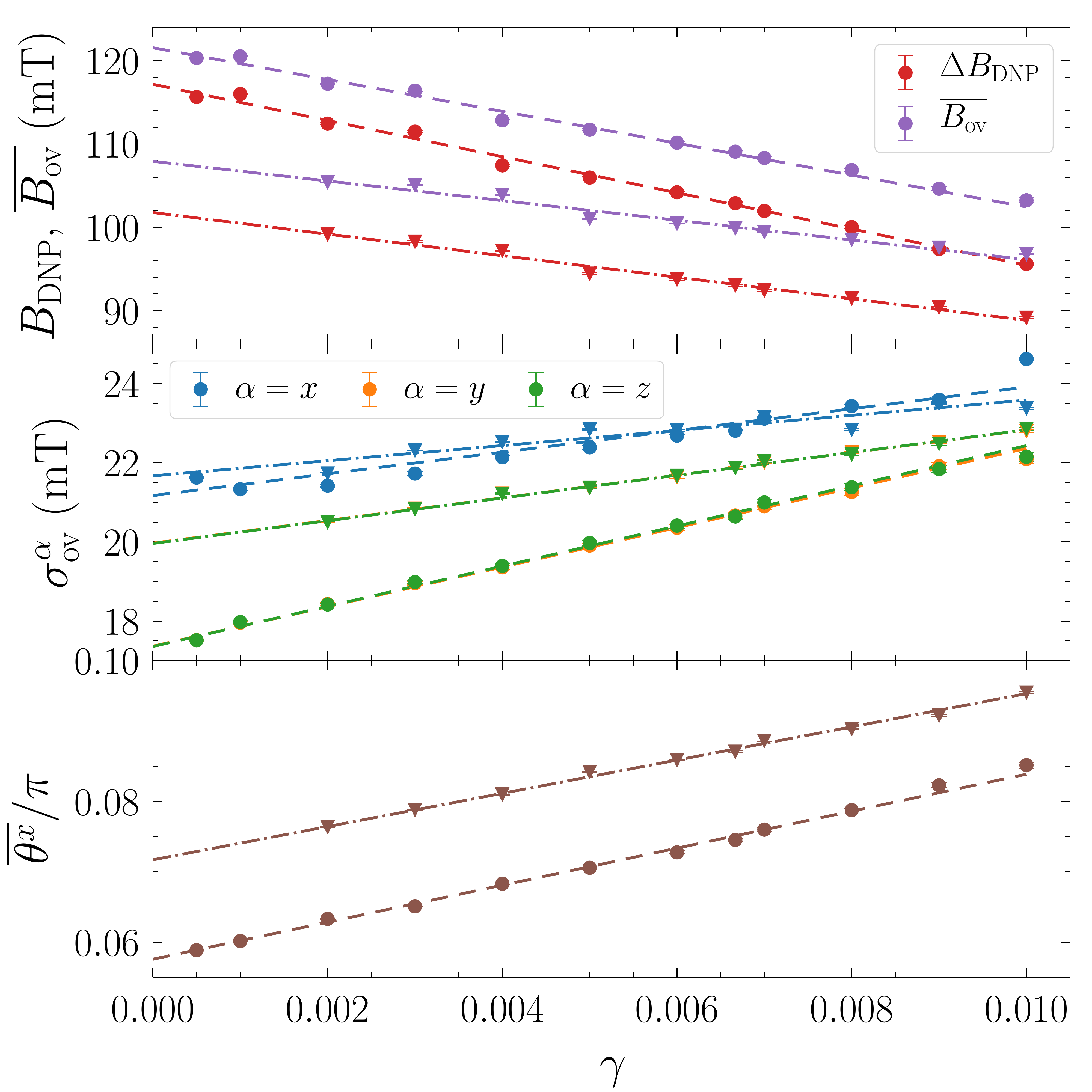}
	\caption{Extended model~II: Limiting values of the DNP $\Bdnp$, the average Overhauser field length $\overline{B_\mathrm{Ov}}$, the Overhauser field standard deviation $\sigma^\alpha_\mathrm{Ov}$, and the average angle $\overline{\theta^x}$, calculated by applying a fit of type~\eqref{eq:arctan} to the data for $\Bext = 0.5$~T (circles) and $1$~T (triangles) for various values of $\gamma$. The fit errors are usually too small to be discernible. The dashed ($\Bext = 0.5$~T) and dash-dotted ($\Bext = 1$~T) lines represent linear fits enabling the extrapolation to an infinite bath size~($\gamma \to 0$).
	}	\label{fig:TWAfT_DNP_extrapolation}
\end{figure}

What is the influence of the system parameters, especially of the effective bath size $\Neff \approx 2/\gamma$, on the DNP behavior?
First, we find that the rate of DNP scales linearly with $\gamma$ and the data for $\Bext = 0.5$~T and $1$~T strongly suggests a $\Bext^{-2}$ dependence.
Note that these are the same scaling laws as for NIFF~\cite{scher18}.

For a more detailed analysis of the influence of the bath size, we apply fits of type~\eqref{eq:arctan} to the data for $\Bext = 0.5$~T (circles) and $1$~T (triangles) for various values of $\gamma$ and plot the resulting saturation values in Fig.~\ref{fig:TWAfT_DNP_extrapolation}.
It turns out that the dependence on $\gamma$ is linear for all observables, hence, a linear fit enables the extrapolation $\gamma \to 0$, i.e., the limit of interest for QDs with $10^4 - 10^6$ nuclear spins.
For $\Bdnp$, this extrapolation yields a value of $117$~mT for $\Bext = 0.5$~T and a value of $102$~mT for $\Bext = 1$~T. 
For ${\Bext = 0.5}$~T, the standard deviations $\sigma^y_\mathrm{Ov}$ and $\sigma^z_\mathrm{Ov}$ are reduced by about $40\%$ from their initial value of about $29$~mT to only $17.4$~mT.
The standard deviation $\sigma^x_\mathrm{Ov}$ decreases slightly less to $21.2$~mT.
For $\Bext = 1$~T, this anisotropy is less pronounced with limiting standard deviations of $\sigma^x_\mathrm{Ov} = 21.7$~mT and $\sigma^y_\mathrm{Ov} = \sigma^z_\mathrm{Ov} = 20$~mT.

Comparing the results in the limit $\gamma \to 0$ with the data for finite values of $\gamma$, we conclude that all main effects are already present for, e.g., $\gamma = 0.01$. 
The size of the spin bath influences only the precise values for the observables, but their order of magnitude turns out to be independent of $\gamma$.

Let us briefly discuss the role of the dephasing time $\Tnstar$ on DNP.
According to the definition~\eqref{eq:Overhauser_variance}, the typical fluctuation strength of the Overhauser field is proportional to $(\Tnstar)^{-1}$, hence, we also expect this dependence for the maximum DNP.
For QDs with ${\gamma = 0.01}$ at $\Bext = 0.5$~T, we find a DNP of $\Bdnp = 127$~mT for $\Tnstar = 0.5$~ns, $96$~mT for $\Tnstar = 1$~ns, and $45$~mT for $\Tnstar = 2$~ns.
For $\Bext = 1$~T, the dependence is similar, i.e., a larger dephasing time $\Tnstar$ corresponds to a smaller DNP. 
We cannot confirm the $(\Tnstar)^{-1}$ dependence in our data with certainty, but it fits sufficiently well to provide an educated guess.
For $\gamma \to 0$ and $\Tnstar \approx 4$~ns, which corresponds to the QD sample discussed in Ref.~\cite{fischer18}, we estimate a DNP of about $30$~mT based on the scaling with $(\Tnstar)^{-1}$.
A direct simulation of this particular sample is out of reach because the required computational effort scales roughly with $(T_\mathrm{n}^*)^3$.
In any case, the DNP is expected to be significantly larger than the root mean square of the Overhauser field distribution.
It would be very interesting to verify or falsify these predictions experimentally.

\subsection{Discussion}

The inclusion of the trion pseudospin dynamics is a crucial step towards the correct description of the underlying pump-probe experiments.
It turns out to have an important qualitative influence on the interplay of SML and NIFF.
In Sec.~\ref{sec:EM_I}, we found a broad range of magnetic fields for which pronounced peaks in the probability distribution $p(\Beff)$ appear at positions fulfilling the ORC.
When including the trion pseudospin dynamics, the majority of magnetic fields $\Bext$ reveal peaks at the ERC, with the only exception being the very narrow but apparently robust feature around $\Bext = 7.8$~T where the ORC emerges.
Around the broad minimum at $\Bext = 3.9$~T, we find very weak frequency focusing in the Overhauser field (see Fig.~\ref{fig:TWAfT_Overhauser}), which is almost independent of the effective bath size. 
In contrast, the frequency focusing without inclusion of the trion pseudospin dynamics is much more pronounced in the vicinity of this magnetic field (see Fig.~\ref{fig:TWA_Overhauser}), but fulfilling the ORC instead of the ERC.
As a result, we find an increase of the revival amplitude in comparison to the SML regime without NIFF due to the constructive interplay of NIFF and SML. 
In contrast, the revival amplitude is strongly suppressed for the EM~I of Sec.~\ref{sec:EM_I}.

In the quantum mechanical model with $N=6$ nuclear spins studied by \citet{klein18}, odd resonances emerge in the vicinity the NRC~\eqref{eq:NRC} for $\Bext = 3.9$~T ($k=1$) with an accompanied minimum of the revival amplitude. 
They also find a minimum around $\Bext = 7.8$~T ($k=2$), but it is much broader than in our simulations.
At this particular value neither the ERC nor the ORC is fulfilled in their model, similar to what we find in Fig.~\ref{fig:TD_Overhauser} for our initial model.
But the trion pseudospin dynamics was not accounted for in Ref.~\cite{klein18}.
We expect that its inclusion would also have a significant influence on the results of the quantum mechanical model.

Why do even instead of odd resonances appear around $\Bext = 3.9$~T upon inclusion of the full trion pseudospin dynamics? 
One can think of the additional precession of the trion pseudospin around the effective magnetic field as a strong perturbation, especially when the external magnetic field~$\Bext$ is large.
Effectively, this leads to a decoupling of the trion from the ground state such that our equations of motion become very similar to those studied by \citet{glazo12a}, where only the ERC appears.
Note that in the derivation of the ORC~\eqref{eq:ORC}~\cite{jasch17}, the trion only follows an exponential decay as described by Eq.~\eqref{eq:eom_Jz}.
Hence, the ORC~\eqref{eq:ORC} in its original form does not hold anymore because the trion dynamics becomes more complex when accounting for its full dynamics as in Eq.~\eqref{eq:eom_fT}.
At present, however, the origin of the persistent sharp minimum at $\Bext = 7.8$~T remains unclear.

The experimental situation on this issue is unclear, but the tools for a systematic study are available.
By applying a radio frequency field, \citet{evers18} are able to scramble the nuclear spins in the QDs such that they do not contribute to the revival amplitude by means of NIFF. 
By applying this approach to a broad range of magnetic fields, a systematic comparison between SML without NIFF and SML with NIFF is possible. 
Such an experimental study is likely to clarify whether or not NIFF always leads to an increase of the revival amplitude.
Moreover, a detailed search for sharp features in the vicinity of the NRC~\eqref{eq:NRC}, especially for $k=2$, is promising.
Note that multiple NRCs are possible when considering all individual magnetic moments of the isotopes of the QD sample instead of an average one.
The consideration of several isotopes is likely to lead to a more complex structure of the revival amplitude as a function of the magnetic field strength. 
This issue is beyond the scope of the present work, but will be the topic of future research.

The finding of a DNP of about $120$~mT in the simulation is interesting because it implies a certain increase of the coherence time due to the associated narrowing of the Overhauser field distribution.
But for a substantial increase, a much larger DNP is required~\cite{urba13,chekhovich13}. 
Yet, the DNP found in our simulations is significantly larger than the root mean square of the Overhauser field distribution.
Indications for DNP in similar but simpler simulations can be seen in the results of Ref.~\cite{scher18}, but the effect remained unnoticed.

Note that the nondeterministic pulse is not responsible for DNP.
We also find DNP when we use the deterministic pulse model~\eqref{eq:pulse} in combination with the full trion dynamics.
For this combination, DNP as a function of the magnetic field shows a fairly similar behavior as depicted in Fig.~\ref{fig:TWAfT_DNP}.
However, almost perfect NIFF fulfilling the ERC with ${\Slim \to 0.5}$ appears when studying this particular combination.
Only for $\Bext = 7.8$~T, the ORC with $\Slim \to -1/6$ emerges instead of the ERC.
But since the broad minimum around $\Bext = 3.9$~T is missing and NIFF is almost perfect (unlike in the experiments~\cite{jasch17,klein18,evers18,klein18}), we do not study this combination in more detail.

As argued above, the model studied by \citet{glazo12a} is similar to our model because the trion pseudospin effectively decouples from the ground state for large magnetic fields.
In their model, DNP is predicted analytically by studying the stability of the fixed point given by the ERC~\eqref{eq:ERC}. 
Without any additional nuclear spin relaxation, the resonance condition turns out to be an unstable fixed point so that DNP is possible.
In agreement with Ref.~\cite{glazo12a}, changing the helicity of the pulses does not change the DNP direction.
The emergence of DNP could be less efficient in the experiments due to weak nuclear spin relaxation~\cite{glazo12a}.

Experimental hints for DNP in this type of experiment already exist.
In the measurements of Ref.~\cite{jasch17}, the distribution of Larmor frequencies, extracted from the measured real-time evolution via pump-probe spectroscopy, is shown, and one can discern a shift from the bare Larmor frequency resulting from the external magnetic field, possibly due to DNP.

Since about two orders of magnitude more pulses are required to reach the saturation of the DNP in comparison to the steady state of the revival amplitude due to NIFF, it is well possible that the DNP steady state is not reached in the experiments so far.
Experimentally, it takes about a minute to reach a strong revival amplitude for a magnetic field of $6$~T~\cite{greil07a}.
By applying the suggested scaling with $\Bext^{-2}$, which has yet to be confirmed experimentally, we estimate a strong DNP to emerge within $5$~minutes for a magnetic field of $1$~T and within $3$~hours for a magnetic field of $6$~T.
A systematic experimental study of this interesting feature is definitely called for.

\section{Extended model III:\\Inhomogeneous ensemble of quantum~dots}
\label{sec:EM_III}

In the results of the previous section, beats appear in the initial dephasing of the signal $S^z(t) - J^z(t)$ (Fig.~\ref{fig:TWAfT_S_timeevolution}), which decay during the trion lifetime $\tau_0 = 400$~ps. 
However, these beats vanish noticeably quicker in measurements~\cite{greil06a}.
Moreover, the ensemble dephasing time $T_2^*$ shows a strong magnetic field dependence in the experiments~\cite{greil06a,greil06b,fischer18}, which cannot be explained by the Overhauser field.

\begin{figure}[t!]
	\centering
	\subfloat{\includegraphics[width=\columnwidth]{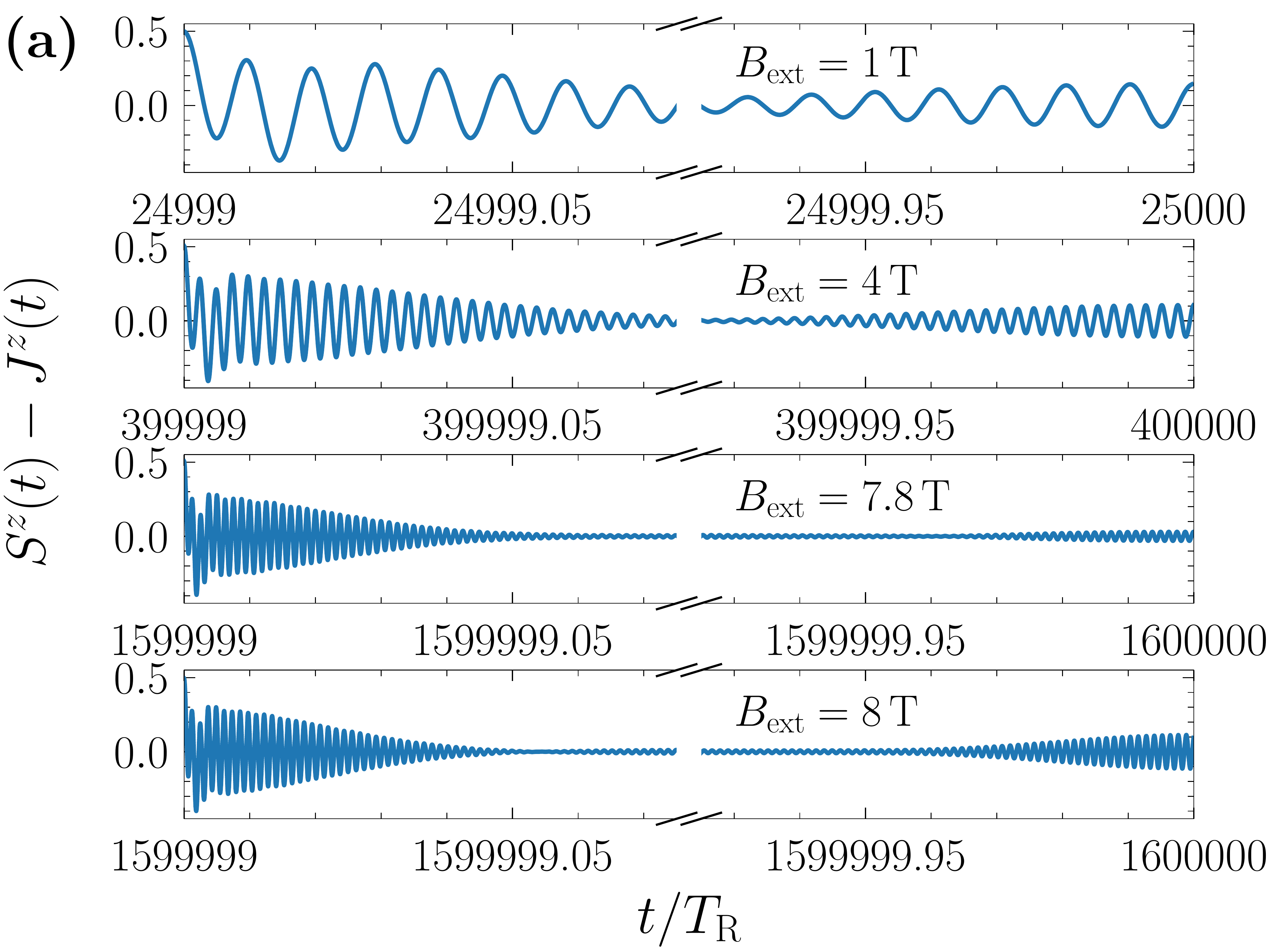} \label{fig:TWAfT_gspread_S_timeevolution}} \\
	\subfloat{\includegraphics[width=\columnwidth]{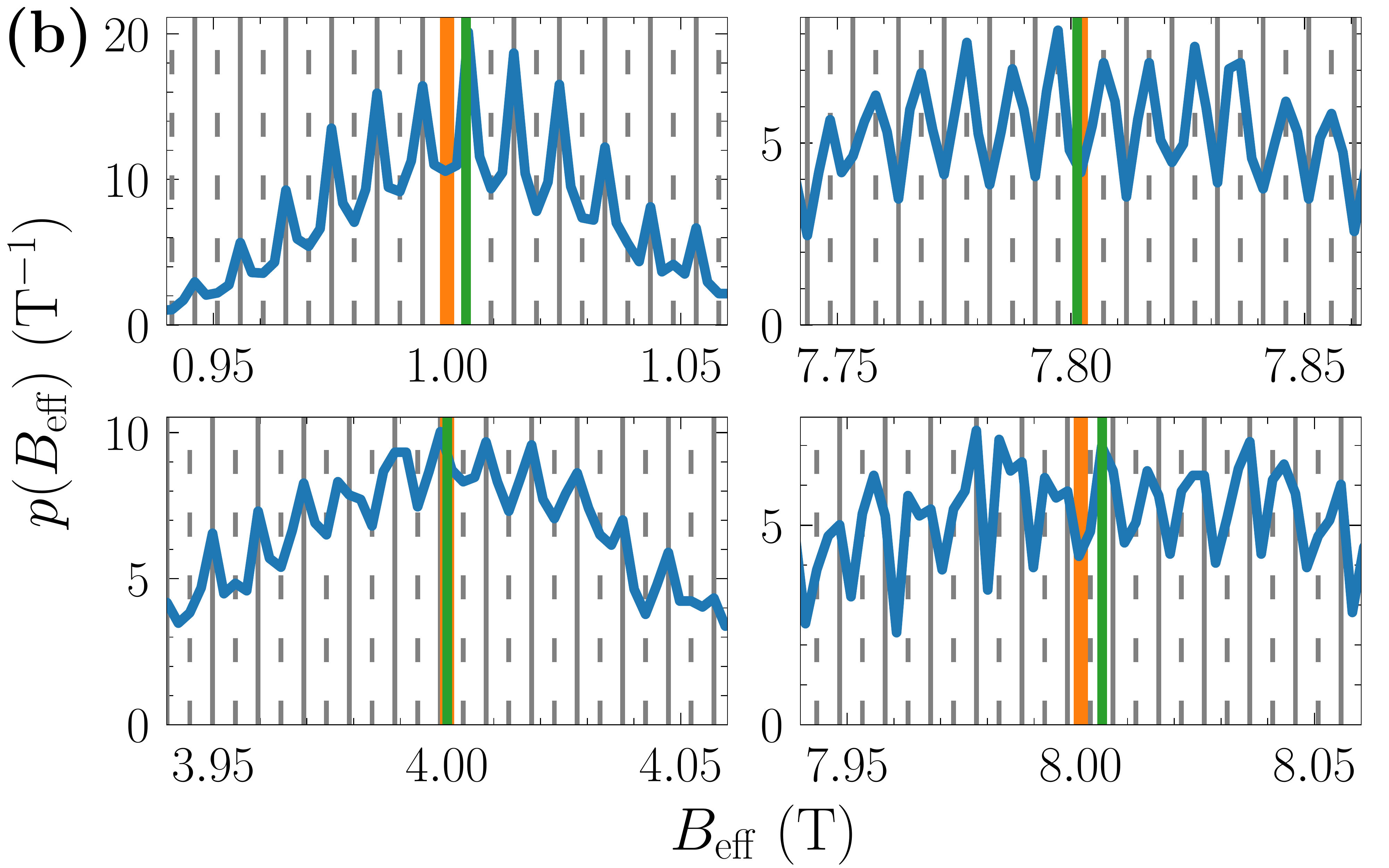} \label{fig:TWAft_gspread_Overhauser}} \\
	\caption{Numerical results for the extended model~III (EM~III) introduced in Sec.~\ref{sec:EM_III}: 
		(a) Spin dynamics between consecutive pulses after a long train of pulses for various external magnetic fields $\Bext$ for an inhomogeneous ensemble of QDs. 			
		(b) Probability distributions of the effective magnetic field $p(\Beff)$ when the revival amplitude is in the saturation regime for various external magnetic fields $\Bext$ (orange vertical lines). The gray solid and dashed vertical lines represent the values of $\Beff$ fulfilling the ERC~\eqref{eq:ERC} and~\eqref{eq:ORC}, respectively. The green vertical line represents the mean value of the distribution. 
		Parameters: $\gamma = 0.004$, $\Tnstar = 1$~ns, $\ge = 0.555$, $\Delta \ge = 0.005$, $\gh = 0.15$, $\Delta \gh = 0.05$.
	}
	\label{fig:EM_III}
\end{figure}

Until now, we considered a homogeneous ensemble of QDs with a dephasing time $\Tnstar = 1$~ns.
This is, however, a simplification.
The $g$ factor of the localized electron spin varies slightly from QD to QD because they are not identical, leading to a faster dephasing for large magnetic fields.
We consider resonant optical pumping in this paper, i.e., the $g$ factor of the electron spin in each QD can be modeled by a normal distribution with expectation value $\ge$ and standard deviation $\Delta \ge = 0.005$~\cite{greil06b,fischer18}, leading to the ensemble dephasing time $\T2star$ defined by
\begin{align}
	(\T2star)^{-2} = (\Tnstar)^{-2} + (T^*_\mathrm{inh})^{-2}, \label{eq:T2star}
\end{align}
with the dephasing time ${(T_\mathrm{inh}^*)^{-1} = \Delta \ge \muB \Bext/\sqrt{2}}$ due to the inhomogeneities of the QD ensemble.
This total dephasing time decreases for large magnetic fields while its upper bound is given by $\Tnstar$ for $\Bext \to 0$.

We apply the same modeling to the $g$ factor of the trion pseudospin with standard deviation $\Delta \gh = 0.05$~\cite{yugov07} to account for the fast vanishing of the beats in the time evolution observed in the experiments.
When including a finite spread $\Delta \gh$ for the $g$ factor of the trion pseudospin, the polarization of the ensemble dephases on a timescale which can be shorter than the radiative lifetime $\tau_0$.
The average magnetic moment of the nuclei does not change, i.e., it is still chosen as $\gn \muB  = \ge \muB / 800$.

The implementation of the spread of the $g$ factors is straightforward in our simulations.
It is realized by sampling the $g$ factors from the aforementioned normal distribution around their expectation values given by $\ge = 0.555$ and $\gh = 0.15$.
Results for a different in-plane orientation of the QD sample with $\gh = 0.05$ are presented in Appendix~\ref{app:gh_comparison}, but they show only slight quantitative differences.

Let us discuss the differences to the results of the previous section~(EM~II) when accounting for an inhomogeneous ensemble of QDs~(EM~III).
Figure~\ref{fig:TWAfT_gspread_S_timeevolution} shows the overall faster dephasing for larger magnetic fields $\Bext$, as expected from Eq.~\eqref{eq:T2star}. 
The beats also vanish much quicker, which is in better agreement with the actual experiments~\cite{greil06a,greil06b,greil07c}. 
An even better agreement can be achieved by explicitly fitting the system parameters to experimental results, e.g., the $g$ factor of the trion pseudospin $\gh$, but this is not the goal of this paper.

Instead, we are interested if and how an inhomogeneous ensemble of QDs alters the interplay of SML and NIFF.
Obviously, modeling the $g$ factor of the electron spin by a normal distribution leads to a broadening of the distribution of the effective magnetic field $p(\Beff)$ for large magnetic fields as demonstrated in Fig.~\ref{fig:TWAft_gspread_Overhauser}, but the width of each individual peak due to frequency focusing does not change noticeably.

Without the spread of the electron $g$ factor, peaks also appear in the probability distribution of the mere Overhauser field~$p(B_\mathrm{Ov}^x)$ due to NIFF. They are slightly shifted from the expected resonance positions and they are also slightly broader compared to the probability distribution of the effective magnetic field $p(\Beff)$~\cite{scher18}.
In contrast, for the inhomogeneous ensemble of QDs under study even a rather small magnetic field of $0.5$~T is enough to smear out the resonances in the distribution $p(B_\mathrm{Ov}^x)$.
The minimal width of the peaks is limited by the spread of the $g$ factor $\Delta\ge$ of the localized electron spin. Once this width is larger than the distance $2\pi/\TR$ between adjacent resonances, i.e., $\TR^{-1} \lesssim \Delta \ge \mu_B \Bext$, no peaks can be discerned.
Hence, we find no peaks in the mere Overhauser field distribution $p(B_\mathrm{Ov}^x)$ here, only the distribution of the effective magnetic field $p(\Beff)$ shows a comb like structure.

\begin{figure}[t!]
	\centering
	\includegraphics[width=\columnwidth]{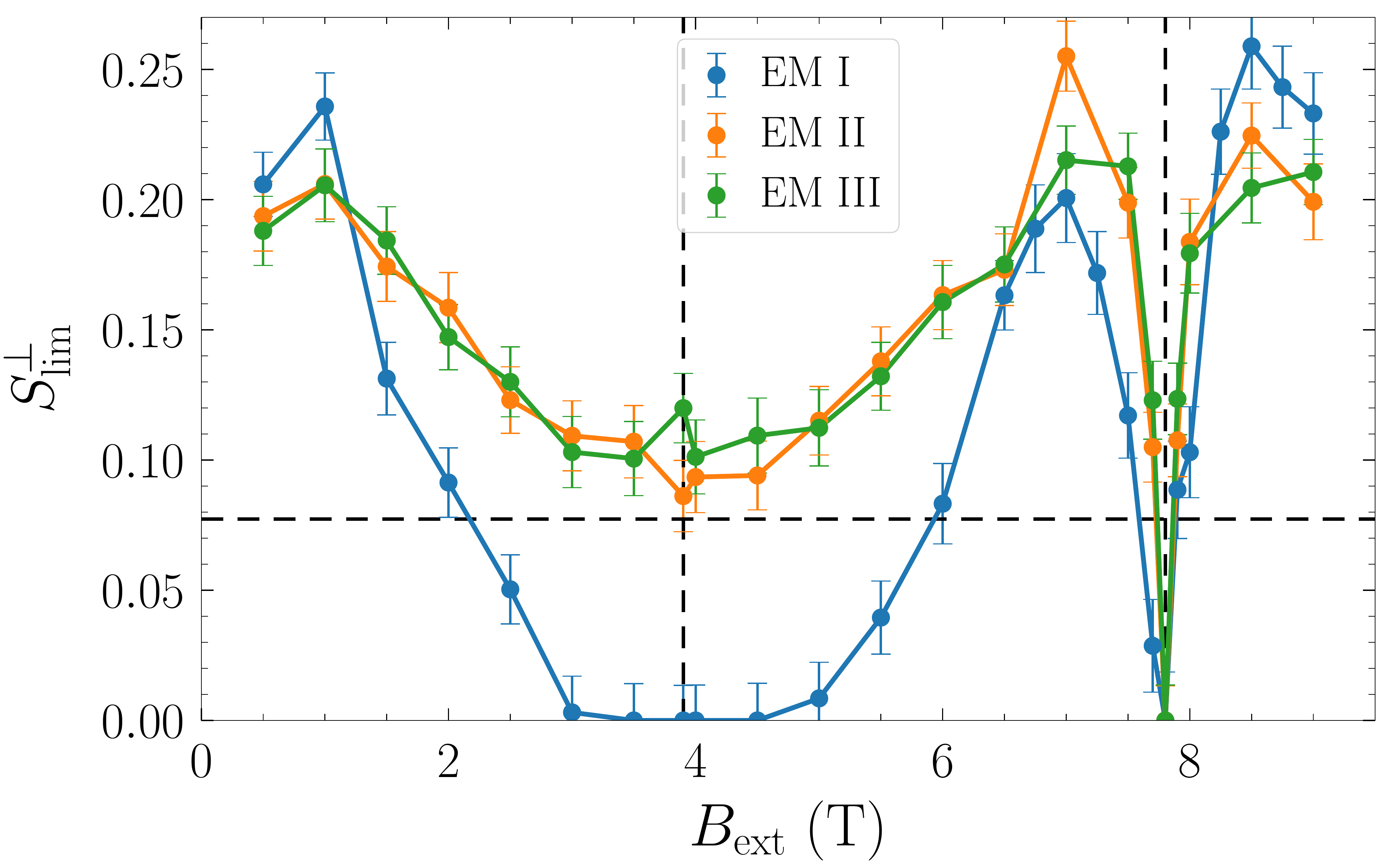}
	\caption{Limiting values $S^\perp_\mathrm{lim}$ of the revival amplitude as a function of the external magnetic field $\Bext$ in the limit of an infinite bath size ($\gamma \to 0$) for the extended models (EM) I, II, and III. The vertical dashed lines represent the NRC~\eqref{eq:NRC} for $k = 1$~and~$2$, the horizontal dashed line indicates the SML steady statue value~$\SSML$.
	}	\label{fig:Slim_Bext_allEM}
\end{figure}

In Fig.~\ref{fig:Slim_Bext_allEM}, we compare the final results for the magnetic field dependence of the revival amplitude $\Slim(\Bext)$ of the EMs I, II, and III in the limit of an infinite effective bath size ($\gamma \to 0$). 
It turns out that no significant difference can be found between EM~II (homogeneous) and III (inhomogeneous),
i.e., the qualitative interplay of SML and NIFF does not change upon inclusion of a finite spread for the $g$ factors.
From the comparison we conclude that the minimum at $\Bext = 7.8$~T is even narrower for the EMs II and III in comparison to EM~I. 
Possibly, this narrow feature remains unmeasured in experiments when the discretization of the magnetic field is chosen too large; we only find it when simulating closely around $\Bext = 7.8$~T.
For this reason, a systematic experimental search for such narrow features would be very interesting.

DNP also occurs even when studying an inhomogeneous ensemble of QDs (see Appendix~\ref{app:gh_comparison}), with an almost identical DNP behavior as in Fig.~\ref{fig:TWAfT_DNP}.
This is expected because the additional small variances of the $g$ factors barely change the dynamics from QD to QD, i.e., the qualitative physics remains the same as in the EM~II. 
The same argument holds for the strong similarity of the results for the EMs~II and~III in Fig.~\ref{fig:Slim_Bext_allEM}.

However, the influence of DNP on the total dephasing time $\T2star$ is smaller for an inhomogeneous ensemble of QDs.
DNP only implies a decrease of the dephasing time $\Tnstar$, which is determined by the strength of Overhauser field fluctuations.
The inhomogeneous dephasing time $T_\mathrm{inh}^*$ is not altered by DNP.
Hence, the relative influence of the narrowed Overhauser field distribution due to DNP on the total dephasing time $\T2star$ as defined by Eq.~\eqref{eq:T2star} is diminished, especially for large magnetic fields.

\section{Conclusion}
\label{sec:conclusion}

We developed an improved semiclassical model for the spin mode locking~(SML) effect in combination with nuclei-induced frequency focusing~(NIFF) in QDs which yields an improved numerical description of various experimental results.
The final model is the result of a combination of several key points while exploiting various scaling arguments.

First, we combined an established semiclassical pulse model often used to describe the excitation of a trion~\cite{yugov09,jasch17}
with an efficient approach to the spin dynamics of the Overhauser field~\cite{fauseweh17,scher18}.
However, the results do not match our expectations gained from a quantum mechanical description of the problem, and they also disagree with the experimental results~\cite{jasch17,klein18}.

Consequently, we improved the pulse model via a nondeterministic description~(EM~I) in which we interpret the pulse as a measurement in order to reduce the discrepancy to quantum mechanical results while being able to cope with large nuclear spin baths.
This step led to considerably improved results which are in qualitative agreement with what is found in the quantum model of Ref.~\cite{klein18}, where only a small bath consisting of $N=6$ nuclear spins is studied.
In this improved model and in agreement with Ref.~\cite{klein18}, both even and odd resonances are found in the probability distribution of the effective magnetic field due to NIFF for different strengths of the external magnetic field.
Importantly, the two kinds of resonances do not appear simultaneously and the corresponding peaks are rather broad due to the mimicked quantum fluctuations.
The emergence of odd resonances in the distribution of the effective magnetic field leads to a reduction of the SML effect in comparison to the case without NIFF.
This means that NIFF leads to a reduction of the revival amplitude in this model for a broad range of magnetic fields.

We improved our theory further by including the full dynamics of the trion pseudospin, resulting in the extended model~II~(EM~II).
We found that the fast Larmor precession of the trion pseudospin acts as a perturbation which suppresses the odd resonances such that the observed behavior of NIFF as a function of the external magnetic field is qualitatively different from the one of the previous model~(EM~I).
In the EM~II, NIFF acts only constructively except for a very narrow regime around a resonance condition for the nuclear spins where their Larmor period between consecutive pulses matches the pulse repetition time.
Even though the $g$ factor of the unpaired heavy-hole spin of the negatively charged trion depends strongly on the in-plane orientation of the QD sample~\cite{yugov07,crooker10}, we find only small quantitative differences between the results.

Furthermore, we observed the emergence of dynamic nuclear polarization~(DNP) of the order of $100$~mT, i.e., the formation of a nonzero average polarization of the nuclear spin ensemble, which can be significantly larger than the typical fluctuations of the Overhauser field.
It is caused by the alignment of the nuclear spins along the axis of the external magnetic field.
Similar behavior can be inferred from the experimental results presented in Ref.~\cite{jasch17} where the spectrum of Larmor frequencies of the localized electron spins is studied.
Importantly, the saturation of the DNP takes about two orders of magnitude longer than the saturation of the revival amplitude due to NIFF.
Its emergence leads to a slight narrowing of the Overhauser field distribution by about one third and thus, also to a slight increase of the dephasing time.
Moreover, we find a similar dependence of the DNP on external magnetic field as for the NIFF.
The absolute value of the DNP is minimal in the vicinity of the nuclear resonance conditions where the nuclear Larmor period corresponds to a multiple of the half pulse reptition time.
For a typical experiment, we estimate the maximum DNP to be $30$~mT for a magnetic field of $1$~T, which is be reached after about 5 minutes.
This DNP is significantly larger than the average fluctuation of the Overhauser field.

Accounting for an inhomogeneous ensemble of QDs led to the extended model~III~(EM~III).
This extension in combination with the full trion pseudospin dynamics (introduced in the EM~II) is crucial for a correct description of the measured spin dynamics between two consecutive pulses.
It does not lead to a qualitatively different DNP behavior, and also not to a different interplay of SML and NIFF.

In all three extended models, the peaks in the Overhauser field distribution are fairly broad compared to, e.g., the initial model discussed in Sec.~\ref{sec:initial_model}.
This is similar to what is found for the quantum mechanical model of Ref.~\cite{klein18}.
Thus, we attribute this behavior to quantum fluctuations captured by the randomness of the pulse model introduced in the EM~I. 

Note that the peak widths are not determined by some additional relaxation time induced by further interactions such as the quadrupolar interaction or dipole-dipole interaction. Their finite width is intrinsic to the studied model at hand.
But, indeed, such further interactions are another possible mechanism which could hinder the efficiency of NIFF.
Moreover, weak nuclear spin relaxation due to such interactions could lead to a reduced DNP efficiency.

The qualitative behavior of the system can be reproduced in a rather simple model. The essential ingredients are the hyperfine interaction of the electron spin with the nuclear spin bath (for which a simple box model is sufficient to achieve a good description), the nondeterministic semiclassical pulse description to take important quantum mechanical corrections into account, and the precession of the trion pseudospin around the external magnetic field which acts as a perturbation to the recombination dynamics.
This model should be realizable on a full quantum level, and the simplification of using a box model might help to overcome the issue of the small bath size.

Further extensions of the model are conceivable.
First, the model should be extended to account for the various isotopes in InGaAs QDs so that several resonance conditions for the nuclear spins act in a combined way.
We expect that this extension leads to a more complex structure in the magnetic field dependence of the revival amplitude, similar to the experimental results of Ref.~\cite{klein18}.

In the present study, we have focused only on the resonant excitation of a trion. 
However, the applied pulse model can be easily generalized to detuned pulses~\cite{yugov09}.
Thereby, one can account for the influence of the inhomogeneous broadening of the QD sample on the trion excitation~\cite{carter09,glazov10,varwig14} and explicitly calculate the Faraday rotation and ellipticity~\cite{yugov09}, which show different dependencies on certain parameters~\cite{varwig14}.
Moreover, this step would enable us to simulate two-color pump-probe experiments~\cite{yugov09,glazov10,varwig14}.
Detuned pulses can lead to the emergence of different resonances as demonstrated in Refs.~\cite{evers18,kopteva19}.
There, the influence of a positive and negative detuning on NIFF is discussed, but a different model is used to describe for the spin dynamics.
In general, the optical Stark effect induced by detuned pulses appears to be very important to accurately describe DNP~\cite{zhukov18b,markmann19,dominguez20}.

A third relevant aspect is the inclusion of a finite pulse duration, which leads to a reduced efficiency of the pulse for large magnetic fields and can also lead to phase shifts for the resonances~\cite{spatzek11,klein18}.

From the experimental side, several clarifications can stimulate progress in understanding the relevant physics.
First, different models suggest different scaling laws for the rate of NIFF as a function of the magnetic field; in our case the rate is reduced by $\Bext^{-2}$.
Second, a systematic comparison of the revival amplitude as a function of the magnetic field for the cases \emph{with} and \emph{without} NIFF is helpful, accompanied by an analysis of the Larmor frequency spectrum with respect to the class of resonances.
Such an experimental study can also reveal the influence of the pulse duration on spin mode locking \emph{without} NIFF by comparing the measured revival amplitude with the analytically obtained steady-state value $\SSML$. This value is independent of the external magnetic field in our model, but a reduced pulse efficiency for large magnetic fields due to a finite pulse duration could be revealed by the suggested experiment.
\citet{evers18} demonstrated that such experiments are realizable by applying an appropriate radiofrequency field to the system, which hinders the frequency focusing of the nuclei.
But so far, measurements at various strengths of the magnetic field have not been carried out.
In this context, studying a potential influence of the in-plane orientation of the QD sample is a further interesting point.
Moreover, the emergence of DNP in the system, for which some evidence in the experimental data in Ref.~\cite{jasch17} exists, is another subject calling for further investigation.

In conclusion, the improved description of the spin dynamics in QDs can help to achieve a better coherent manipulation of this quantum degree of freedom.
This is a prerequisite for applications of QD systems in quantum information and quantum sensing.
Hence, this promising route of research needs to be pursued further.

\acknowledgments
We thank Eiko~Evers, Mikhail~M.~Glazov, Alex~Greilich, Iris~Kleinjohann, and Dmitry~S.~Smirnov for helpful discussions.
We also thank Iris~Kleinjohann for providing the quantum mechanical reference data.
The authors gratefully acknowledge the Gauss Centre for Supercomputing e.V. for supporting this project by providing computing time on the GCS supercomputers Hazel Hen and HAWK at H\"ochstleistungsrechenzentrum Stuttgart~(HLRS).
We also gratefully acknowledge the computing time provided on the HPC cluster LiDO3 at TU Dortmund University, partially funded by the German Research Foundation~(DFG) (Project No. 271512359).
This study has been supported financially by the German Research Foundation~(DFG) and the Russian Foundation for Basic Research~(RFBR) (Grant No. 19-52-12038) in the International Collaborative Research Centre TRR~160 (Projects No.~A4 and No.~A7).

\appendix
\section{Overhauser field dynamics}
\label{app:overhauser}

\subsection{Standard semiclassical approach}

The Overhauser field is defined as the weighted sum of all nuclear spins $\Bov = \sum_{k=1}^{N} A_k \bm I_k$, with $\bm I_k$ being the $k$th nuclear spin, $N$ the number of nuclear spins, and $A_k$ the hyperfine coupling constants defined by~\eqref{eq:couplings}.
In a semiclassical TWA approach~\cite{polkovnikov10,stanek14} to the Overhauser field dynamics, each nuclear spin follows the classical equation of motion
\begin{align}
	\ddt \bm I_k = \left(A_k \S + \gn \mun \Bext \ex \right) \times \bm I_k,
	\label{eq:Ik}
\end{align}
$k \in \{1,2,\dots,N\}$, while the initial conditions are chosen randomly according to normal distributions with expectation value zero and variance $I(I+1)/3$ for each component $I_k^\alpha$~\cite{chen07,stanek14}.
As a result, the initial Overhauser field also follows a normal distribution with expectation value zero and variance 
\begin{align}
	\mathrm{Var}[B_\mathrm{Ov}^\alpha] = \frac{I(I+1)}{3} \AQ^2 =: \frac{2}{(\Tnstar)^2},
	\label{eq:Overhauser_variance}
\end{align}
with $\AQ^2 \coloneqq \sum_{k=1}^N A_k^2$, which defines the dephasing time $\Tnstar$ of the electron spin due to the hyperfine interaction with the nuclear spins with spin $I$. 
In our case, we focus on GaAs QDs so that we have $I = 3/2$.
When studying InGaAs QDs, one needs to also account for ${I = 9/2}$ of the indium isotopes.
This requires a slightly more complicated definition of the variance which includes the relative abundances of the isotopes in a QD.

\subsection{Spectral density approach}
\label{app:SDA}

Since QDs consist of $N = 10^4 - 10^6$ nuclear spins~\cite{merku02,lee05,urba13}, the numerical treatment of the $N$ equations of motion~\eqref{eq:Ik} is unfeasible.
In this paper, we resort to the more efficient spectral density~(SD) approach to calculate the dynamics of the Overhauser field consisting of an infinite number of nuclear spins for a realistic (nonuniform) distribution of the hyperfine couplings.
It was first introduced in Ref.~\cite{fauseweh17} and applied successfully for the description of QDs subjected to periodic pulses in Refs.~\cite{scher18,klein18}.

The SD approach allows us to study an infinite spin bath while the number of \emph{effectively} coupled nuclear spins is finite and given by $\Neff \approx 2/\gamma$~\cite{fauseweh17,scher18,roehr18}.
Instead of calculating the time evolution of each nuclear spin $\bm I_k$ individually, we consider $N_\mathrm{tr} = \mathcal{O}(75)$ auxiliary vectors $\Q_k$ which evolve in time according to the equation of motion 
\begin{align}
	\ddt \Q_k = \left(\epsilon_k  \S + \gn \mun \Bext \ex \right) \times \Q_k,
\end{align}
$k \in \{1,2,\dots,\Ntr\}$, where $\epsilon_k$ is an effective coupling constant derived via application of the SD approach (see below).

In the derivation of this approach~\cite{fauseweh17}, the hyperfine interaction strength of the nuclear spins described by the exponential parametrization~\eqref{eq:couplings} is represented by the linear spectral density $W(\epsilon) = (\epsilon/\gamma) \Theta(\sqrt{2\gamma} \AQ - \epsilon)$, where $\Theta(\epsilon)$ is the Heaviside function.
Note that this approach only works appropriately for ${\gamma \lesssim 0.02}$~\cite{fauseweh17}. 
Otherwise, the assumption of a continuous spectral density is not well justified.

The spectral density is discretized according to the following procedure.
First, we divide the energy range $[0,\,\sqrt{2\gamma} \AQ]$ into $\Ntr$ intervals $\mathcal I_k \coloneqq [\tilde{\epsilon}_{k+1},\,\tilde{\epsilon}_k]$ with
\begin{align}
	\tilde{\epsilon}_k = \lambda^k \sqrt{2\gamma} \AQ \frac{\Ntr - k}{\Ntr}, 
	\quad k \in \{0, 1, 2, \dots, \Ntr\}.
\end{align}
The intervals become exponentially small for increasing~$k$, which is the most efficient choice~\cite{fauseweh17}.

The prefactor $\lambda$ is determined from the relation
\begin{align}
	\lambda = \left( \frac{\Ntr}{\sqrt{2\gamma} \AQ t_\mathrm{max}} \right)^{1/(\Ntr - 1)},
\end{align}
where $t_\mathrm{max} = \np \TR$ is the total simulation time.
For $\lambda = 1$, the discretization is simply equidistant.
It turns out that a good choice for the number of intervals $\Ntr$ is obtained when ensuring that $\lambda \approx 0.87$ holds for long simulations.
For short simulations, we use a minimal number of $\Ntr = 44$ auxiliary vectors $\Q_k$ to ensure a minimal discretization density which still captures the correct physics.

Each interval has the weight
\begin{align}
 W_k \coloneqq \int_{\tilde{\epsilon}_k}^{\tilde{\epsilon}_{k-1}} W(\epsilon) d\epsilon
\end{align}
and the corresponding coupling strength $\epsilon_k$ is given by the average energy
\begin{align}
	\epsilon_k \coloneqq \frac{1}{W_k} \int_{\tilde{\epsilon}_k}^{\tilde{\epsilon}_{k-1}} \epsilon W(\epsilon) d\epsilon, \quad k \in \{1, 2, \dots, \Ntr\}.
\end{align}
The $\Ntr$ auxiliary vectors $\Q_k$ represent the sums of the nuclear spins whose couplings lie within the interval $\mathcal{I}_k$.
Due to the central limit theorem, each initial component of the vectors $\Q_k$ can be drawn from a normal distribution because they represent large linear sums of the nuclear spins, and they are uncorrelated for different $k$.
Thus, we can initialize the $3\Ntr$ components $Q_k^\alpha$, $\alpha \in \{x,y,z\}$, according to normal distributions around zero with variance 
\begin{align}
	\mathrm{Var}[Q_k^\alpha] = \frac{I(I+1)}{3}.
\end{align}
Finally, the Overhauser field is given by the weighted summation
\begin{align}
	\Bov = \sum_{i=1}^{\Ntr} \sqrt{W_k} \Q_k \label{eq:Overhauser_SDA_Appendix},
\end{align}
which leads to the same variance as required by Eq.~\eqref{eq:Overhauser_variance}.

\section{Alternative nondeterministic pulse descriptions}
\label{app:pulse_alternatives}

Establishing a valid nondeterministic semiclassical pulse description, which \emph{on average} keeps the properties of the deterministic pulse model~\eqref{eq:pulse}, is not straightforward.
In this appendix, we discuss several alternatives to the nondeterministic pulse description~\eqref{eq:pulse_distribution} introduced in Sec.~\ref{sec:EM_I} and benchmark them against the deterministic semiclassical pulse~\eqref{eq:pulse} and its quantum mechanical realization used in Ref.~\cite{klein18} in the SML regime without NIFF.
Note that in each approach, the relations~\eqref{eq:pulse_Jz} and~\eqref{eq:pulse_Jxy} for the trion pseudospin~$\J_\mathrm{a}$ remain unchanged.

\subsection{Discrete truncated Wigner approximation}
\label{app:DTWA}

As a first alternative, we apply the \emph{discrete} truncated Wigner approximation (DTWA)~\cite{schachenmayer15} to the deterministic pulse~\eqref{eq:pulse}.
This phase space method only acts on a discrete phase space, which, in turn, gives rise to certain benefits.

In this approach, we sample each spin component $S^\alpha$ from the discrete phase space $\{+1/2,\,-1/2\}$ so that \emph{all} quantum mechanical moments of the spin of the same component are taken into account correctly. 
Moreover, the spin length after a pulse is always given by ${|\S_\mathrm{a}| = \sqrt{3}/2}$.

The ensuring discrete distribution function is defined by
\begin{subequations}
\begin{align}
	P\left(S^\alpha_\mathrm{a} = +\frac{1}{2}\right) &= \frac{1}{2} + \mathrm{E}[S^\alpha_\mathrm{a}],\\
	P\left(S^\alpha_\mathrm{a} = -\frac{1}{2}\right) &= \frac{1}{2} - \mathrm{E}[S^\alpha_\mathrm{a}],
\end{align}
\label{eq:DTWA}%
\end{subequations}
$\alpha \in \{x,y,z\}$, where $\mathrm{E}[S^\alpha_\mathrm{a}]$ is the mathematical expectation value of this probability distribution, which is given by Eq.~\eqref{eq:pulse_distribution}.

This approach works well as long as $|S^\alpha| \le 1/2$, e.g., for the first pulse.
But since the spin with initial length $\sqrt{3}/2$ precesses according to the equation of motion~\eqref{eq:eom_S}, this condition does not necessarily hold for every pulse, leading to the appearance of negative probabilities in Eq.~\eqref{eq:DTWA}.
Our heuristic solution consists of effectively truncating the probability distribution, i.e., we set $P(S^z_\mathrm{a} = 1/2) = 1$ when $S^z_\mathrm{b} > 1/2$.
However, this alters the resulting expectation value of the distribution and thereby, also the SML steady state.
Another drawback of the DTWA is the broken rotational spin symmetry because certain spin axes are treated in a special way.

\subsection{Trion probability approach}
\label{app:TP}

In this approach, we use $S^z_\mathrm{b}$ to determine the probability for the excitation of a trion. In this interpretation, the system realizes either the ground state electron spin $\S$ \emph{or} the trion pseudospin $\J$ directly after the pulse.

The circularly polarized laser pulse $\sigma^-$ only excites the electron spin if it is in the state $\ket{\downarrow}$.
In the classical representation, this means that ${S^z_\mathrm{b} = -1/2}$ leads to the excitation of a trion, thus $S^z_\mathrm{a} = 0$ and $J^z_\mathrm{a} = -1/2$.
For ${S^z_\mathrm{b} = +1/2}$, no trion is excited so that $S^z_\mathrm{a} = +1/2$ and $J^z_\mathrm{a} = 0$ follows.
More general, the probability to find the spin in the state $\ket{\downarrow}$ and therefore to excite a trion is given by
\begin{align}
	P_\downarrow = \frac{1}{2} - S^z_\mathrm{b}. \label{eq:TPA}%
\end{align}
If no trion is excited, the $z$~component of the electron spin simply takes the value $S^z_\mathrm{a} = +1/2$ while the $x$ and $y$ component are sampled from a normal distribution with expectation value zero and variance $1/4$ to account for the second moment of spin-$1/2$ operators.
Alternatively, sampling from the discrete phase space introduced in Appendix~\ref{app:DTWA} is possible, but the results are worse.

Mathematically, this procedure leads to expectation values which are identical to the deterministic pulse relation~\eqref{eq:pulse}.
However, the same issue as discussed in Appendix~\ref{app:DTWA} arises. 
Since negative probabilities can appear, the probability distribution requires needs to be truncated,
i.e., we set $P_\downarrow = 1$ if $S^z_\mathrm{b} < -1/2$ and $P_\downarrow = 0$ if $S^z_\mathrm{b} > 1/2$. 
Eventually, this leads to a deviation of the expectation value from~\eqref{eq:pulse_distribution} and accordingly, also to a deviation from the SML steady state.

A possible solution consists of scaling the spin $\S_\mathrm{b}$ to the Bloch sphere of spin length $1/2$ before applying the pulse. We will see, however, that this procedure leads to the emergence of an unwanted phase shift.

\subsection{Comparison to established pulse descriptions}
\label{app:pulses_comparison}

Let us compare the various nondeterministic pulse descriptions to the established deterministic pulse relation~\eqref{eq:pulse}~\cite{yugov09} and its quantum mechanical realization used in Ref.~\cite{klein18} in the SML regime.
Note that we do not include the trion pseudospin dynamics here because it has no relevant influence on the SML regime.

\begin{figure}[t!]
	\centering
	\includegraphics[width=\columnwidth]{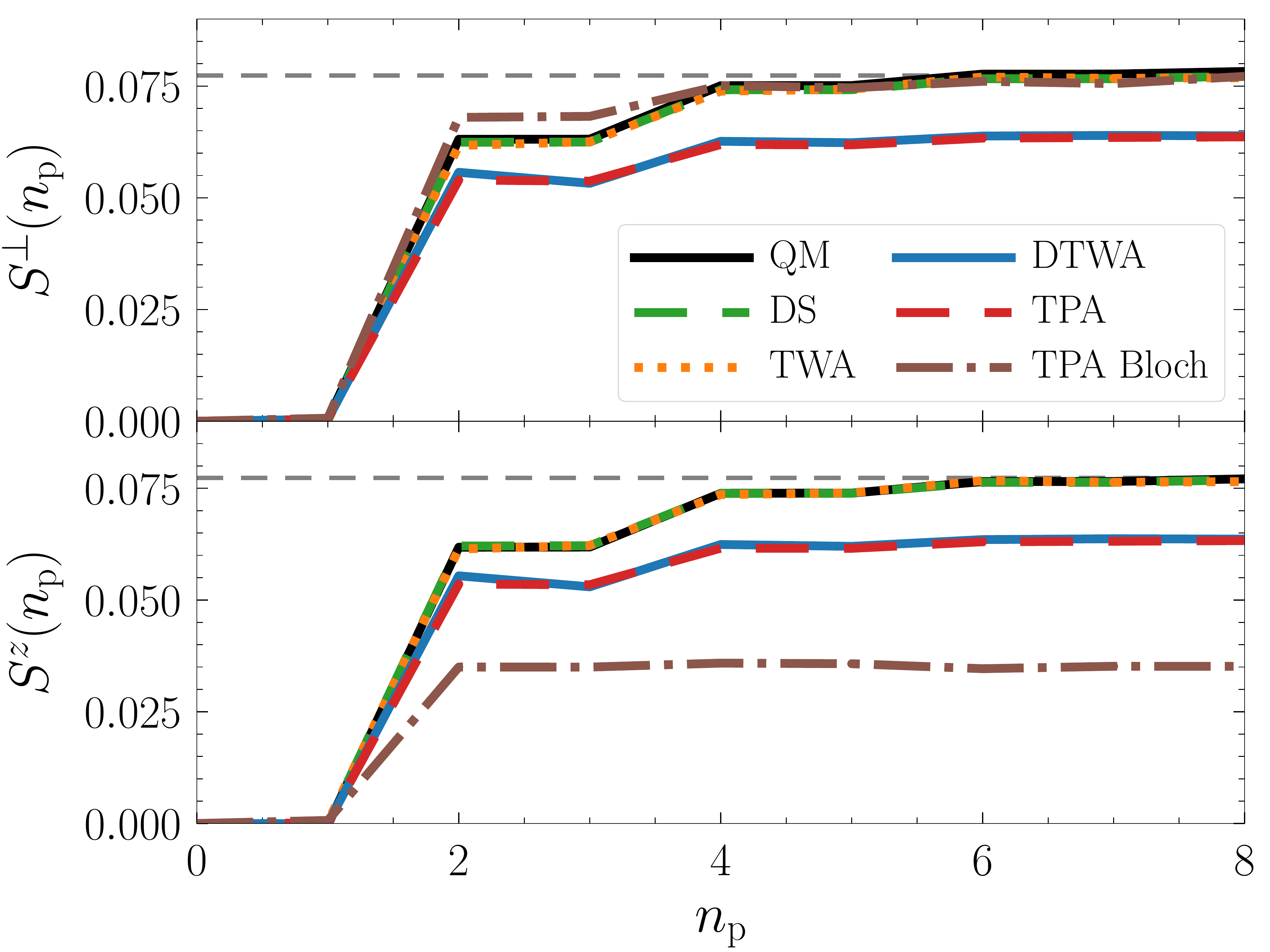}
	\caption{Build-up of the revival amplitude $S^\perp$ (upper panel) and its $z$~component (lower panel) to the SML steady state for the different pulse descriptions discussed in the main text of Appendix~\ref{app:pulse_alternatives}, combined with the equations of motion~\eqref{eq:eom} and~\eqref{eq:Qk} used for the initial model in Sec.~\ref{sec:initial_model}. The gray horizontal dashed line represents the analytical steady state~${\SSML \approx 0.07735}$ [Eq.~\eqref{eq:SML_steadystate}]. Parameters: $\Bext = 1$~T, $\gamma = 0.01$, averaged over $M = 10^6$ independent trajectories.
	}	\label{fig:SML_pulses}
\end{figure}

Figure~\ref{fig:SML_pulses} shows the revival amplitudes $S^\perp$ (upper panel) and the corresponding $z$~components $S^z$ (lower panel) for the following pulse descriptions:
quantum mechanical (QM, black), deterministic semiclassical (DS, green dashed) [Eq.~\eqref{eq:pulse}], TWA [Eq.~\eqref{eq:pulse_distribution}, orange dotted], DTWA [Eq.~\eqref{eq:DTWA}, blue], trion probability approach (TPA) [Eq.~\eqref{eq:TPA}, red dashed], and TPA with scaling to the Bloch sphere (see end of Appendix~\ref{app:TP}, brown dash-dotted).

As expected, the QM and the DS results are almost identical. 
Small deviations stem from the fact that the QM results are obtained for only $N=6$ nuclear spins, which requires an additional ensemble average to get rid of finite size effects.
These results serve as our benchmark. 
They show the expected revival amplitude of the SML steady-state value $\SSML \approx 0.07735$ and there is almost no difference between $S^\perp$ and its $z$~component, i.e., there is no phase shift.

The results for the TWA pulse, which we introduce and apply in the EM~I of Sec.~\ref{sec:EM_I}, are in perfect agreement with the benchmark results (QM and DS). Small deviations stem mainly from the statistical nature of the ensemble average.
We use $M = 10^6$ configurations here to calculate the ensemble average; the statistical deviations are proportional to $1/\sqrt{M}$.

For the remaining nondeterministic pulse description, we find no satisfying agreement with the benchmark results.
As expected, the DTWA and TPA pulse show a too small steady state revival amplitude. 
Interestingly, the results are identical apart from small statistical fluctuations in the SML regime, but the behavior in the NIFF regime is extremely different (not shown).

By scaling the spin vector $\S_\mathrm{b}$ to the Bloch sphere before the application of the TPA, the revival amplitude $S^\perp$ reaches the correct steady state.
However, the revival amplitude is about two times larger than its $z$~component, i.e., a significant phase shift is introduced by scaling to the Bloch sphere. 
Such a phase shift does not appear in the benchmark data.

\section{In-plane orientation of the quantum~dot ensemble}
\label{app:gh_comparison}

The $g$ factor of the unpaired heavy-hole spin $\gh$ of the negatively charged trion $X^-$ strongly depends on the in-plane orientation of the QD sample, with values ranging from $\gh = 0.05$ to $0.15$~\cite{yugov07,crooker10}.
In the extended models (EM)~II and~III, the spin precession of this hole spin around the external magnetic field [see Eq.~\eqref{eq:eom_fT}] acts as a perturbation to the recombination dynamics which is responsible for the ORC~\eqref{eq:ORC}.

\begin{figure}[b!]
	\centering
	\subfloat{\includegraphics[width=\columnwidth]{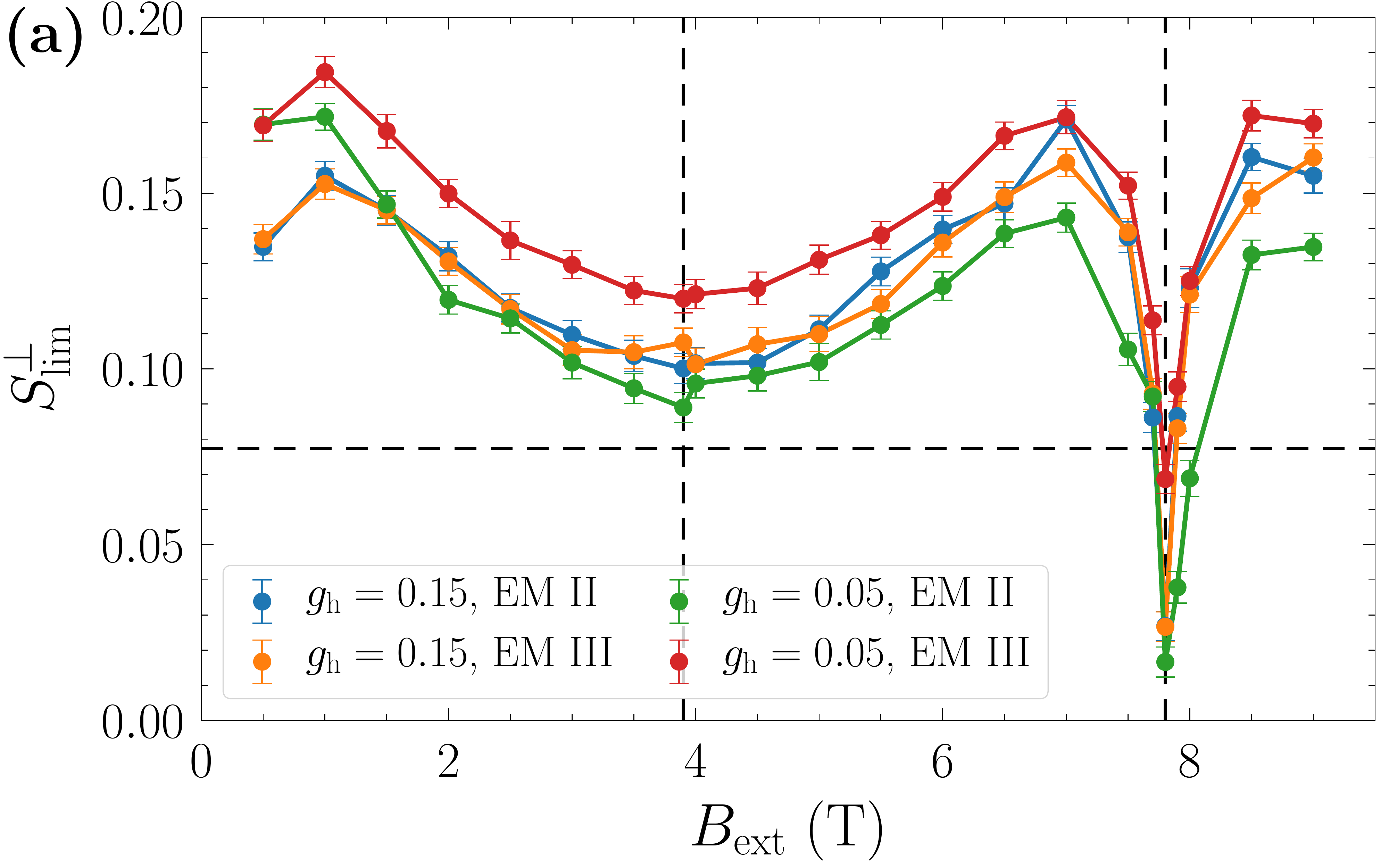} \label{fig:Slim_Bext_gh_comparison}} \\
	\subfloat{\includegraphics[width=\columnwidth]{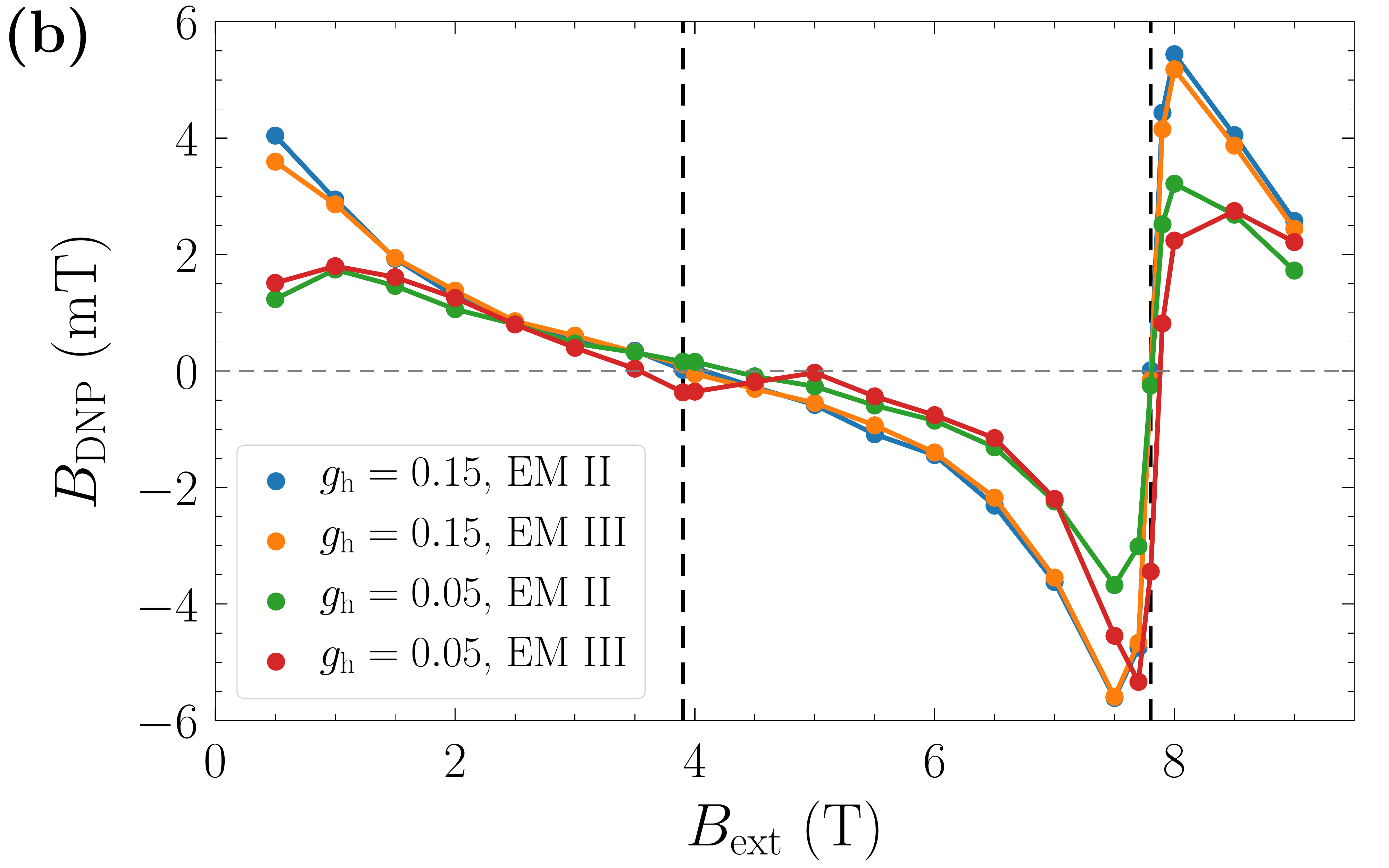} \label{fig:DNP_Bext_gh_comparison}}
	\caption{Influence of the $g$ factor of the trion pseudospin $\gh$ on NIFF and DNP for the extended models~(EM)~II~and~III for $\gamma = 0.004$: (a) Limiting values~$S^\perp_\mathrm{lim}$ of the revival amplitude as a function of the external magnetic field $\Bext$. The vertical dashed lines represent the NRC~\eqref{eq:NRC} for $k = 1$~and~$2$, the horizontal dashed line indicates the SML steady statue value~$\SSML$.
		(b) DNP $\Bdnp$ as a function of the external magnetic field $\Bext$. The number of applied pulses is chosen such that $\Slim$ is approximately in its steady state.
	}
	\label{fig:gh_comparison}
\end{figure}

In Fig.~\ref{fig:Slim_Bext_gh_comparison}, we compare the results for the revival amplitude~$\Slim$ as a function of the external magnetic field~$\Bext$ using $\gamma = 0.004$ for $\gh = 0.15$ and $\gh = 0.05$~(EM~II). 
We also study the influence of a finite spread of the $g$ factors existing in an inhomogeneous quantum dot ensemble~(EM~III).
We choose $\Delta \ge = 0.005$~\cite{greil06b,fischer18} and $\Delta \gh = 0.05$~\cite{yugov07} as in Sec.~\ref{sec:EM_III}.
While the results for $\gh = 0.15$ with~(EM~III, orange) and without~(EM~II, blue) the spread of the $g$ factors are identical within the accuracy of our data, there is a shift of $\Slim$ for $\gh = 0.05$ when comparing the EM~II (green) with~III (red), with larger revival amplitudes when the spread is included. 
The behavior is very similar in the limit of an infinite bath size ($\gamma \to 0$, not shown).
For $\gamma = 0.01$ at $\Bext = 7.8$~T using $\gh = 0.05$, we actually find $\Slim > \SSML$ (not by much), hinting at the emergence of a weak ERC instead of the usual ORC at this particular magnetic field. In the limit $\gamma \to 0$, the revival amplitude is also larger than zero at this particular magnetic field, but it is still smaller than $\SSML$ (ORC).

Comparing $\gh = 0.05$ (red) to $0.15$ (orange) for the EM~III, the revival amplitude is slightly larger for ${\gh = 0.05}$.
Differences between two samples of QD ensembles were reported in Ref.~\cite{klein18}.
In this context, it would be interesting to study a potential influence of the in-plane orientation of the samples on these measurements experimentally.

We compare the complementary results for the DNP $\Bdnp$ as a function of the magnetic field $\Bext$ in Fig.~\ref{fig:DNP_Bext_gh_comparison}.
The DNP reached after $\Slim$ is approximately in saturation is plotted, i.e., $\Bdnp$ is not in its steady state yet as this would require about two orders of magnitude more pulses, which is out of reach for our simulations for the broad range of magnetic fields.
Again, for $\gh = 0.15$ no significant differences are found between the EMs~II and~III.
For $\gh = 0.05$, the DNP is generally weaker.
This also holds true for $\Bext = 0.5$~T and $1$~T in the DNP saturation regime (not shown).

Interestingly for $\gh = 0.05$ using the EM~III, there is a slight buckling around $\Bext = 3.9$~T, where no DNP is found for the other cases. We also find a rather large DNP at $\Bext = 7.8$~T, where no DNP occurs in the other simulations.
But note that this particular case needs to be treated cautiously because the value $\Delta \gh = 0.05$ was measured for $\gh = 0.15$~\cite{yugov07}.
The behavior probably is related to the fact that we have chosen ${\Delta \gh = \gh = 0.05}$ here, i.e., the $g$ factor can change its sign and it can be very close to zero in many cases.
Especially when it is close to zero, there is no fast Larmor precession of the trion pseudospin around the external magnetic field acting as a perturbation anymore, resulting in a qualitative change of the physics.

%

\end{document}